\documentclass[11pt,twoside]{article}
\usepackage{graphicx,natbib,subfigure}%,epsfig,epstopdf
\usepackage{CS18}
\usepackage{url}
\usepackage{hyperref}
\hypersetup{colorlinks=true}
\hypersetup{citecolor=blue}
\usepackage[all]{hypcap}

\newcommand{\Mjup}{$M_{\mathrm{Jup}}$}
\newcommand{\Msol}{$M_{\odot}$}
\newcommand{\masyr}{$\mathrm{mas}\ \mathrm{yr}^{-1}$}

\begin{document}

\title{RESULTS FROM \emph{BASS}, THE BANYAN ALL-SKY SURVEY}

\author{Jonathan Gagn\'e\altaffilmark{1,2},\, David Lafreni\`ere\altaffilmark{1},\, Ren\'e Doyon\altaffilmark{1},\, Jacqueline K. Faherty\altaffilmark{3,4},\, Lison Malo\altaffilmark{1,5}, \'Etienne Artigau\altaffilmark{1}}
\affil{\altaffilmark{1} D\'epartement de Physique, Universit\'e de Montr\'eal, C.P. 6128 Succ. Centre-ville, Montr\'eal, Qc H3C 3J7, Canada}
\affil{\altaffilmark{2} IPAC Fellow}
\affil{\altaffilmark{3} Department of Terrestrial Magnetism, Carnegie Institution of Washington, Washington, DC 20015, USA}
\affil{\altaffilmark{4} Hubble Fellow}
\affil{\altaffilmark{5} Canada-France-Hawaii Telescope, 65-1238 Mamalahoa Hwy, Kamuela, HI 96743, USA}

\begin{abstract}

We present results from the BANYAN All-Sky Survey (\emph{BASS}), a systematic all-sky survey for brown dwarf candidates in young moving groups. We describe a cross-match of the \emph{2MASS} and \emph{AllWISE} catalogs that provides a list of 98\,970 potential nearby dwarfs with spectral types later than M5 with measurements of proper motion at precisions typically better than 15 \masyr, as well as the Bayesian Analysis for Nearby Young AssociatioNs II tool (BANYAN~II) which we use to build the \emph{BASS} catalog from this \emph{2MASS}--\emph{AllWISE} cross-match, consisting of more than 300 candidate members of young moving groups. We present the first results of a spectroscopic follow-up of those candidates, which allowed us to identify several new low-mass stars and brown dwarfs displaying signs of low gravity. We use the \emph{BASS} catalog to show tentative evidence for mass segregation in AB~Doradus and Argus, and reveal a new $\sim$ 13 \Mjup\ co-moving companion to a young low-mass star in \emph{BASS}. We obtain a moderate-resolution near-infrared spectrum for the companion, which reveals typical signs of youth and a spectral type L4~$\gamma$.

\end{abstract}

\section{INTRODUCTION}

Nearby young moving groups (YMGs) such as TW~Hydrae (TWA; 8 -- 12~Myr; \citealp{1989ApJ...343L..61D}; \citealp{2004ARA&A..42..685Z}), $\beta$~Pictoris ($\beta$PMG; 26 -- 29~Myr; \citealp{2001ApJ...562L..87Z}; \citealp{2014arXiv1406.6750M}), Tucana-Horologium (THA; 20 -- 40~Myr; \citealp{2000AJ....120.1410T}; \citealp{2000ApJ...535..959Z}), Carina (20 -- 40~Myr; \citealp{2008hsf2.book..757T}), Columba (20 -- 40~Myr; \citealp{2008hsf2.book..757T}),  Argus (ARG; 30 -- 50~Myr; \citealp{2000MNRAS.317..289M}) and AB~Doradus (ABDMG; 70 -- 120~Myr; \citealp{2004ApJ...613L..65Z}) provide a unique opportunity for the study of young, age-calibrated very low-mass stars and brown dwarfs (BDs) in the Solar neighborhood. The current census of bona fide members of those groups is still limited to early-type ($\leq$ M0) stars, whereas a small fraction of the expected M0--M5 population has been only recently explored (\citealp{2012ApJ...758...56S}; \citealp{2013ApJ...762...88M}; \citealp{2013ApJ...774..101R}; \citealp{2014AJ....147..146K}). The identification of the even fainter BD members is a challenging task which requires a kinematic and spectroscopic follow-up of a large number of faint targets: only a handful of such BD candidate members have been identified so far (\citealp{2005ApJ...634.1385M}; \citealp{2007ApJ...669L..97L}; \citealp{2010ApJ...715L.165R}; \citealp{2012A&A...548A..26D}; \citealp{2013ApJ...777L..20L}; \citealp{2013AJ....145....2F}; \citealp{2014ApJ...783..121G}; \citealp{2014ApJ...785L..14G}).

The Banyan All-Sky Survey (\emph{BASS}) was initiated to search for this missing $>$ M5 population of young low-mass stars and BDs in nearby moving groups. Completing this population will open the door to answering a number of fundamental questions on the formation process of such low-mass objects, \emph{e.g.} by constraining the low-mass end of the initial mass function and studying the physical properties of these objects in coeval populations of well-defined ages. A few results from the \emph{BASS} survey were already published, including the discovery of a planetary-mass companion to a M5.5 binary low-mass star in THA \citep{2013A&A...553L...5D} and the first L dwarf candidate member to TWA \citep{2014ApJ...785L..14G}.\\

We describe here the candidate selection process that was used to build a list of several hundreds of young, late-type candidate members to moving groups (Sections~\ref{sec:select} and \ref{sec:bass}); the status of our spectroscopic follow-up (Sections~\ref{sec:youth} and \ref{sec:spectro}); as well as a number of preliminary results from the \emph{BASS} survey (\hyperref[sec:res]{Section~\ref*{sec:res}}).

\section{SELECTION OF CANDIDATES}\label{sec:select}

\begin{figure}
	\centering
	\subfigure[\emph{2MASS}-\emph{AllWISE} $\mu_\alpha\cos\delta$ precision]{\includegraphics[width=0.495\textwidth]{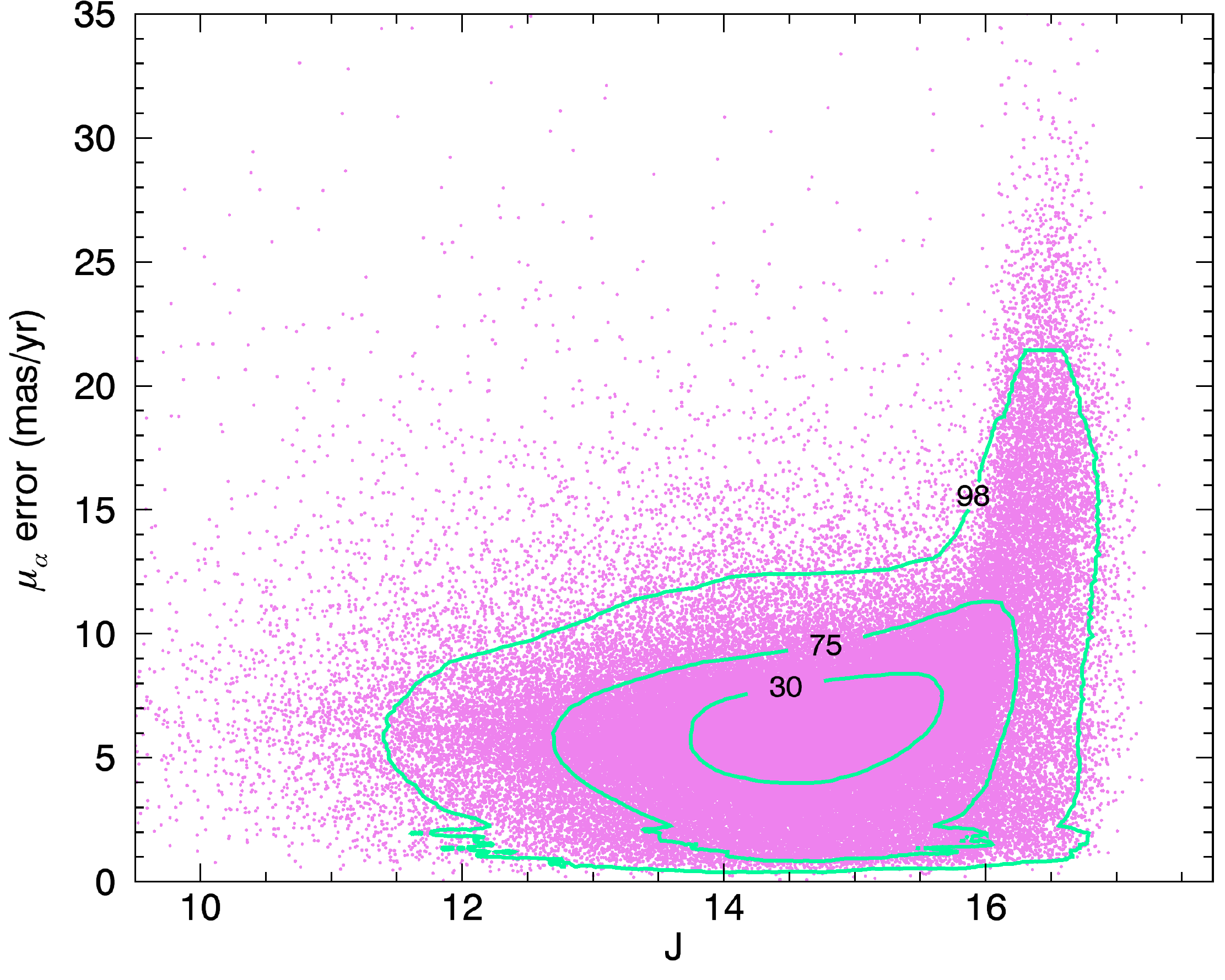}}
	\subfigure[\emph{2MASS}-\emph{AllWISE} $\mu_\delta$ precision]{\includegraphics[width=0.495\textwidth]{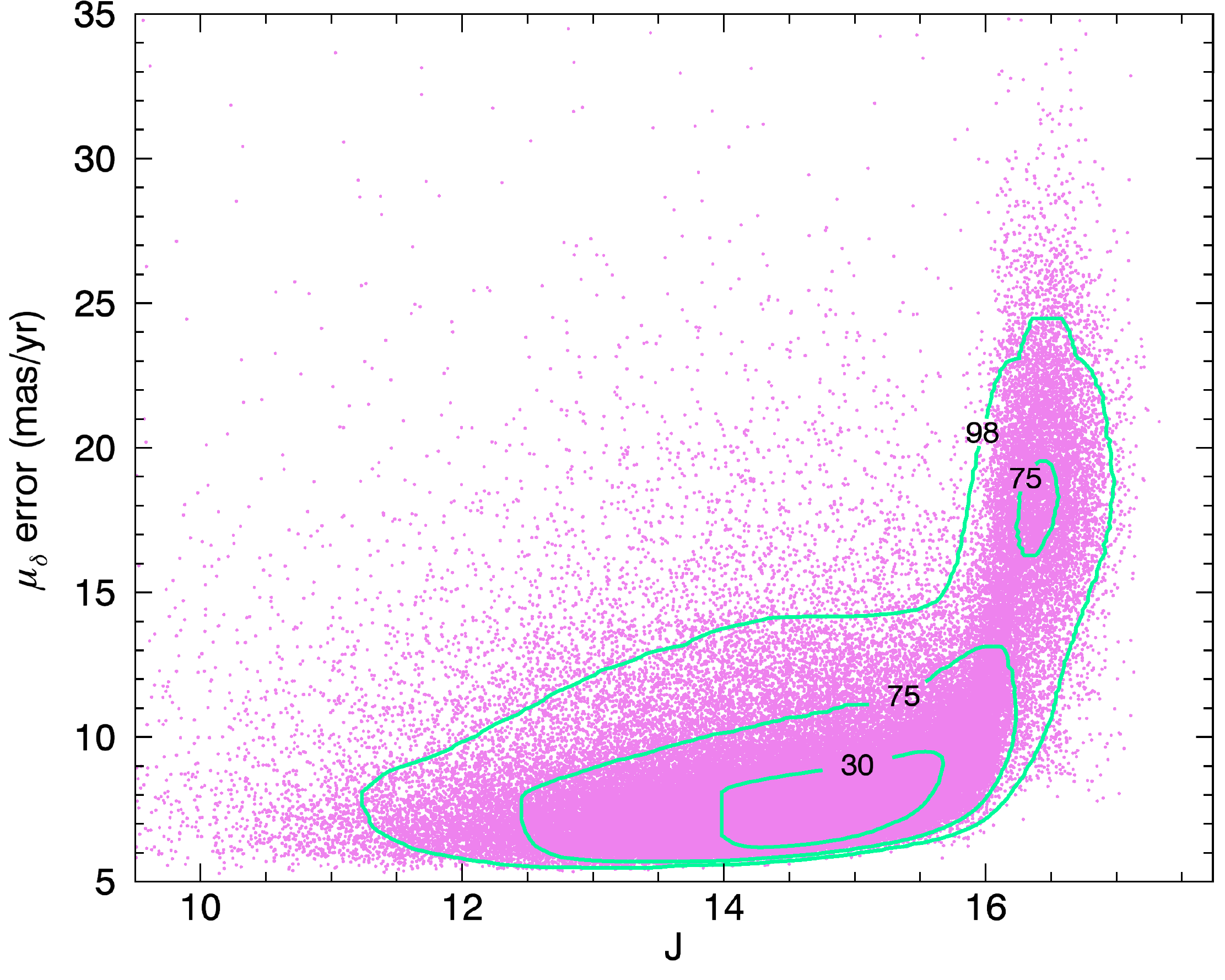}}
	\caption{Precision of \emph{2MASS}--\emph{AllWISE} proper motions derived in this work, as a function of \emph{2MASS} $J$ apparent magnitude (pink dots). Pale green contour lines include 30\%, 75\% and 98\% of our sample, respectively. We obtain typical precisions of 5--15 \masyr\ in the bright regime ($J < 16$) and 5--25 \masyr\ in the faint regime.}
	\label{fig:PM_PREC}
\end{figure}

Candidate members of YMGs in the \emph{BASS} survey were selected from an initial cross-match of the Two Microns All-Sky Survey (\emph{2MASS}; \citealp{2006AJ....131.1163S}) with the \emph{AllWISE} survey (\citealp{2010AJ....140.1868W}; \citealp{2014ApJ...783..122K}). All sources in \emph{AllWISE} were already matched with their \emph{2MASS} counterpart if it is located within an angular distance of 3". We used all unmatched sources in both catalogs to identify additional matches with angular distances up to 25", for all 173\,443 sources outside of the Galactic plane and surviving a set of various quality and color filters. We rejected spurious matches for which colors were not consistent with late-type objects (\emph{i.e.}, $K_S - W1 < 0.153$ or $K_S - W1 > 2$). We used the astrometric measurements in both catalogs to determine the proper motions with typical precisions of 5--15 \masyr\ for bright sources ($J < 16$) and 5--25 \masyr\ for fainter sources 
(\hyperref[fig:PM_PREC]{Figure~\ref*{fig:PM_PREC}}). We then selected all sources with a proper motion larger than 30 \masyr\ at a confidence level larger than 5$\sigma$. This set of 98\,970 objects corresponding to potential nearby, later-than-M5 dwarfs, was used as the input sample for the identification of new candidate members of nearby, young moving groups. Cross-matching this sample with the Initial Gaia Source List (Vizier catalog \emph{I/324/igsl3}) revealed that our \emph{2MASS}--\emph{AllWISE} proper motions are consistent with those of UCAC4 \citep{2013AJ....145...44Z} and PPMXL \citep{2010AJ....139.2440R}, with reduced $\chi^2$ values of 1.27 and 1.03 for $\mu_\alpha\cos\delta$ and $\mu_\delta$, respectively. \\

%Figure : BASS, PM Precision

\begin{figure}
	\centering
	\includegraphics[width=0.99\textwidth]{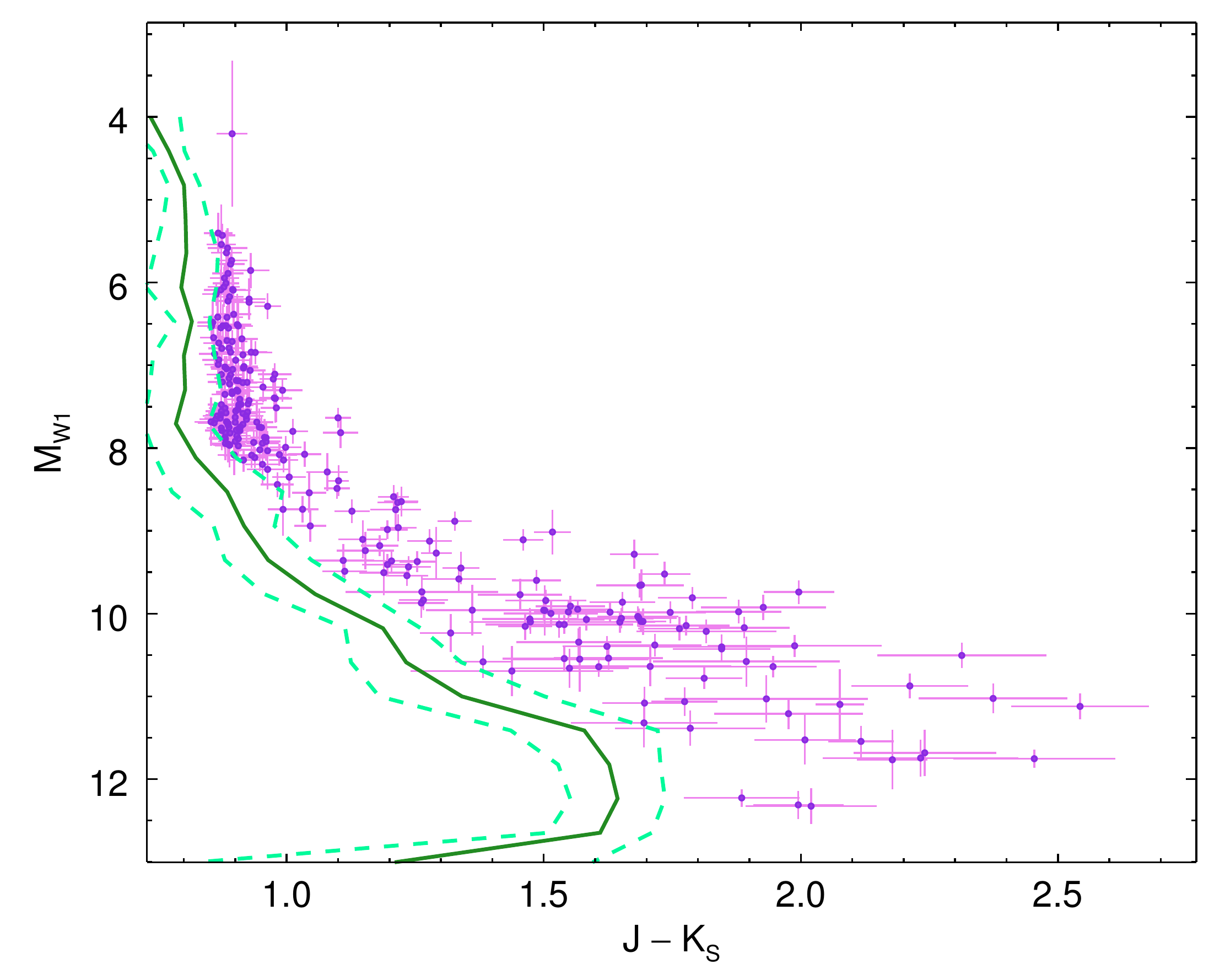}
	\caption{\emph{2MASS} $J-K_S$ color versus \emph{AllWISE} absolute $W1$ magnitude for the field sequence (thick green line) and its scatter (dashed pale green lines), compared to candidate members of YMGs identified in the \emph{BASS} survey (filled purple circles; using their most probable statistical distances from the \href{http://www.astro.umontreal.ca/\textasciitilde gagne/banyanII.php}{BANYAN~II} tool). Objects in \emph{BASS} were selected so that they fall on the red side of the field sequence in this CMD, which is typical of young, inflated (thus brighter) M-type dwarfs, as well as young L-type dwarfs which have thicker dust clouds in their photosphere (causing redder near-infrared colors).}
	\label{fig:JK_CMD}
\end{figure}

\begin{figure}
	\centering
	\includegraphics[width=0.99\textwidth]{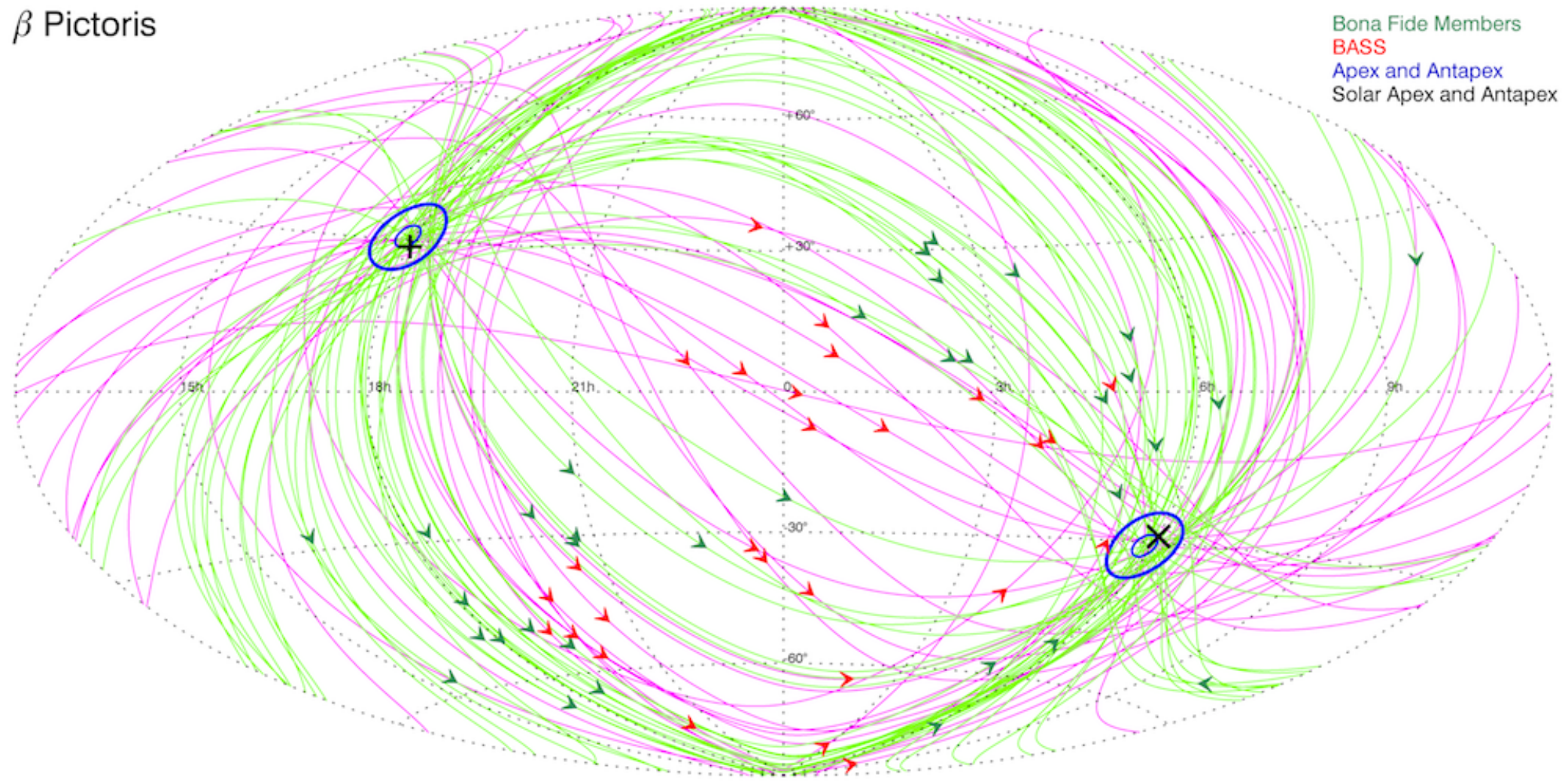}
	\caption{Proper motion and position of $\beta$PMG candidates from the \emph{BASS} survey (red arrows) and their corresponding great circles (pink lines), compared to those of bona fide members of $\beta$PMG (green arrows and lines). It can be seen that bona fide members as well as candidates all point away from the antapex and towards the apex of $\beta$PMG (blue circles), which is known as the property of common proper motion, and is due to the similar space velocities $UVW$ of the members of a moving group projected on the Celestial Sphere. The Solar apex and antapex are displayed as the black cross and plus symbols, respectively.}
	\label{fig:PM_BETAPIC}
\end{figure}

We used the Bayesian Analysis for Nearby Young AssociatioNs~II tool\footnote[1]{Publicly available at \url{http://www.astro.umontreal.ca/\textasciitilde gagne/banyanII.php}.} (\href{http://www.astro.umontreal.ca/\textasciitilde gagne/banyanII.php}{BANYAN~II}; \citealp{2014ApJ...783..121G}) to identify candidate members of YMGs considered here, amongst the input sample of potential nearby $\geq$ M5 dwarfs described above. This tool takes as inputs the sky position, proper motion and \emph{2MASS} $J$, $H$, $K_S$ as well as \emph{AllWISE} $W1$ and $W2$ apparent magnitudes of a given source and compares them to spatial and kinematic models for the Galactic position $XYZ$ and space velocity $UVW$ of these associations and field stars, as well as field and young dwarf sequences in two distinct color-magnitude diagrams (CMDs; $M_{W1}$ versus $J-K_S$ and $M_{W1}$ versus $H-W2$). Young objects are expected to fall on the red side of the field sequence in these CMDs, due to their larger luminosity (a consequence of their inflated radius) and/or the enhanced presence of dust in their photosphere (in the case of L-type BDs; see \hyperref[fig:JK_CMD]{Figure~\ref*{fig:JK_CMD}}). The comparison with these models is performed using a naive Bayesian classifier, where different hypotheses consist in membership to the field or the YMGs considered here. Since radial velocity and distance are not known for the \emph{BASS} catalog, those parameters are marginalized in Bayes' theorem (\emph{i.e.}, a range of radial velocities and distances are tested and the output probability density functions are integrated over those two parameters). The \href{http://www.astro.umontreal.ca/\textasciitilde gagne/banyanII.php}{BANYAN~II} tool outputs a statistical prediction for the distance and radial velocity and a probability associated to each hypothesis. For more information, we refer the reader to \cite{2014ApJ...783..121G} and \href{http://www.astro.umontreal.ca/\textasciitilde gagne/banyanV.php}{J.~Gagn\'e et al. (submitted to ApJ)}. All sources with a Bayesian probability lower than 10\% or with a predicted position on the blue side of the field sequence in any of the two CMDs described above were rejected. All 273 sources with CMD positions at least 1$\sigma$ redder than the field sequence were selected to define the \emph{BASS} catalog (\hyperref[fig:JK_CMD]{Figure~\ref*{fig:JK_CMD}}), whereas the 275 sources with colors within the field scatter were selected to define the low-priority \emph{BASS} (\emph{LP-BASS}) catalog. In \hyperref[fig:PM_BETAPIC]{Figure~\ref*{fig:PM_BETAPIC}}, we show the sky position and proper motion of candidate members of $\beta$PMG in \emph{BASS}, compared to its bona fide members. It can be seen that the proper motions of both samples are consistent and point towards the apex of the moving group.

%Figure : J - K CMD

%Figure : PM Betapic

\section{THE \emph{BASS} CATALOG}\label{sec:bass}

We performed a literature search for all 548 sources in the \emph{BASS} and \emph{LP-BASS} catalogs to gather any relevant information, such as multiplicity, age, radial velocity, trigonometric distance, spectral types or membership to associations, to refine Bayesian membership probabilities with the \href{http://www.astro.umontreal.ca/\textasciitilde gagne/banyanII.php}{BANYAN~II} tool. We also determined an estimated maximal hit rate of 87\% and 74\% in the \emph{BASS} and \emph{LP-BASS} catalogs, respectively, which is obtained by calculating the fraction of candidates that were rejected by any additional information in the literature, compared to the number of candidates for which additional information was consistent with suspected membership.

We retrieve a total of 60/97 of all known candidates or bona fide members to the YMGs considered here in the \emph{BASS} catalog, as well as an additional three in \emph{LP-BASS}. Most (22/36) of the remaining objects were missed because of quality filters that were applied in the construction of our input list, whereas the 14 others were missed because of low Bayesian probabilities.

%Figure : Histogram SPT
\begin{figure}
	\centering
	\includegraphics[width=0.99\textwidth]{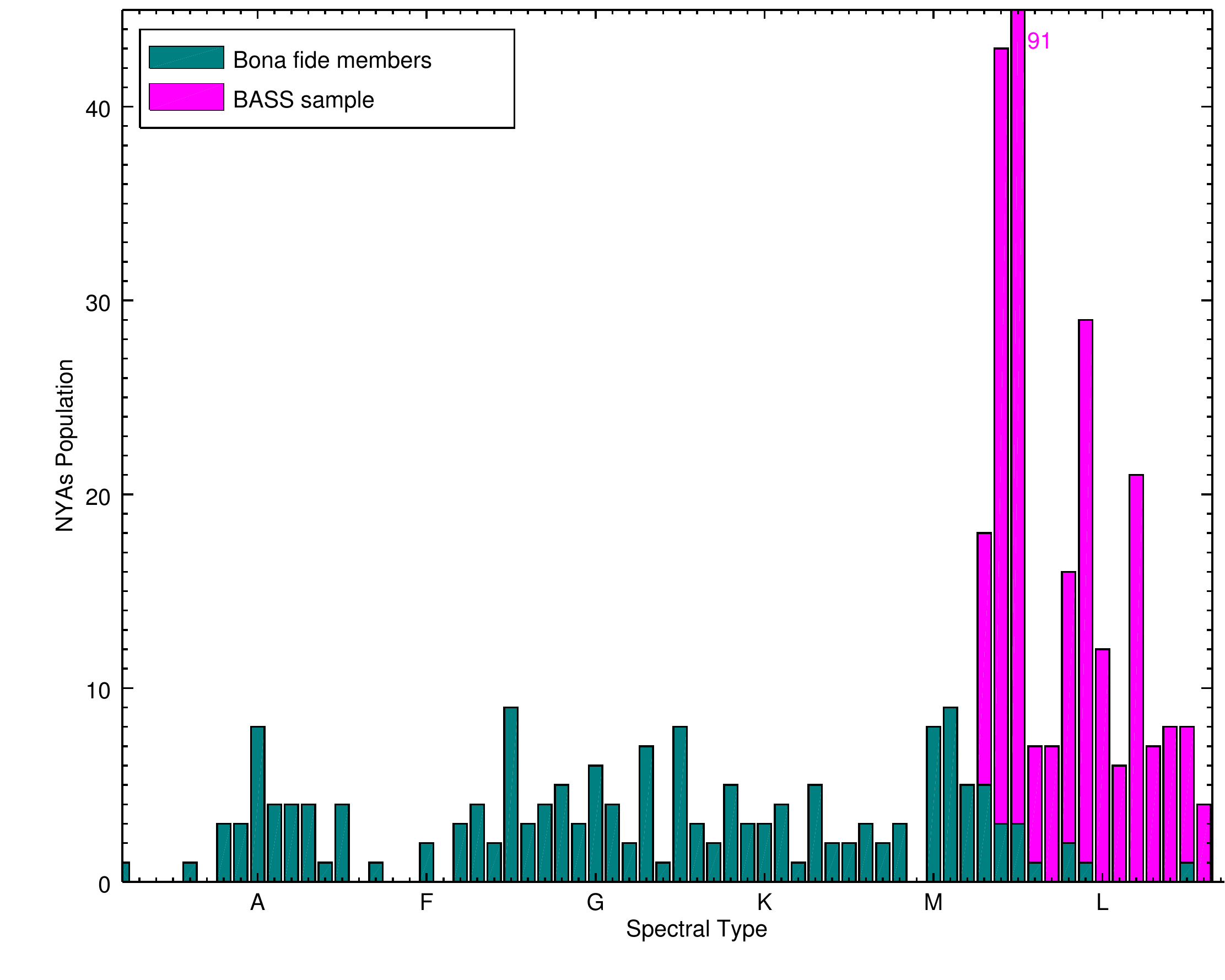}
	\caption{Histogram of estimated spectral types for the \emph{BASS} catalog (pink bars), compared with known bona fide members in the associations considered here (green bars). The vertical range has been truncated for visibility (the bar for spectral type M5 extends to a value of 91). This survey explores a yet poorly known population of very late-type and low-mass young objects in YMGs.}
	\label{fig:SPT_BASS}
\end{figure}

We used the BT-SETTL atmosphere models (\citealp{2013MSAIS..24..128A}; \citealp{2013A&A...556A..15R}), AMES-Cond isochrones \citep{2003A&A...402..701B}, the \emph{2MASS} and \emph{AllWISE} magnitudes, as well as the most probable statistical distances of \emph{BASS} candidates to estimate their spectral types (\hyperref[fig:SPT_BASS]{Figure~\ref*{fig:SPT_BASS}}) and masses. We find that the \emph{BASS} sample potentially contains 101 previously unknown young BDs, 22 of them having estimated masses below 13 \Mjup.

\section{SIGNATURES OF YOUTH IN LOW-MASS STARS AND BROWN DWARFS}\label{sec:youth}

Young low-mass stars and BDs have a larger radius as a consequence of their gravitational contraction phase, which is still ongoing. The thermal energy released by this contraction makes them hotter than field dwarfs of the same mass: both effects conspire to make them more luminous than older objects of a similar mass. Since spectral types are mostly dependent on temperature, young dwarfs are less massive and more inflated than older objects of the same spectral type, and thus have a lower surface gravity. This in turn yields a lower pressure in their atmospheres, decreasing the effects of pressure broadening and collision-induced absorption of the H$_2$ molecule. These effects respectively cause young objects to have a a smaller equivalent width of atomic absorption lines such as K~I, Na~I, FeH and CrH (\citealp{2004MNRAS.355..363L}; \citealp{2004ApJ...600.1020M}; \citealp{2009AJ....137.3345C}), as well as a triangular-shaped continuum in the $H$-band \citep{2010ApJ...715L.165R}. Furthermore, atmospheric cloud thickness is enhanced in young L dwarfs, which causes stronger VO absorption and redder near-infrared (NIR) colors (\citealp{2007ApJ...657..511A}; \citealp{2008ApJ...681..579B}; \citealp{2008ApJ...686..528L}). \\

\cite{2006ApJ...639.1120K} suggest appending a Greek-letter suffix to the spectral type nomenclature, in order to differentiate field dwarfs with a normal surface gravity ($\alpha$) from younger dwarfs with a mild low surface gravity ($\beta$) or those with a very low surface gravity ($\gamma$). \cite{2009AJ....137.3345C} follow this nomenclature and assign a gravity classification to several new young L dwarfs, by visually comparing their optical spectra with those of various spectroscopic standards. \cite{2013ApJ...772...79A} presents a quantitative method to assign a gravity classification to $>$ M5 dwarfs using various spectroscopic indices and equivalent widths suitable for NIR spectroscopy with resolutions $R \geq 75$ or $R \geq 750$ (depending on individual indices). They find that young dwarfs do not display uniform signs of low gravity, in the sense that young objects of the same age and spectral type can reveal their low-gravity nature through distinct properties. For this reason, a set of several low-gravity indices must be considered altogether to assign a low-gravity classification. They build several sequences of spectral types -- spectroscopic indices which are each used to assign a low-gravity score associated to a particular gravity-sensitive index, then all scores are combined together to a final gravity classification of either \emph{field gravity} (FLD-G), \emph{intermediate gravity} (INT-G) or \emph{very low gravity} (VL-G). They report that the FLD-G, INT-G and VL-G classifications are consistent with the respective $\alpha$, $\beta$ and $\gamma$ classifications used in the optical. For this reason, we follow the suggestion of \cite{2006ApJ...639.1120K} and use a Greek-letter classification in the remainder of this work.

\section{SPECTROSCOPIC FOLLOW-UP}\label{sec:spectro}

We initiated a low- to mid-resolution ($R \sim 75$ to $R \sim 6\,000$) spectroscopic follow-up of the \emph{BASS} catalog to assign spectral types and identify signs of youth in the NIR and optical, using GMOS \citep{2004PASP..116..425H} and FLAMINGOS-2 \citep{2004SPIE.5492.1196E} at the Gemini-North and Gemini-South telescopes; SpeX \citep{2003PASP..115..362R} at the IRTF; as well as FIRE \citep{2013PASP..125..270S} and MAGE \citep{2008SPIE.7014E.169M} at the MAGELLAN telescopes. We obtained more than 255 hours of queue observing and 15 nights of classical observing that were dedicated to this project in the 2012A to 2014B semesters.

%Figure : Candidates, batch 1
\begin{figure}
	\centering
	\subfigure[\emph{BASS}~YL001 (L0~$\beta$)]{\includegraphics[width=0.495\textwidth]{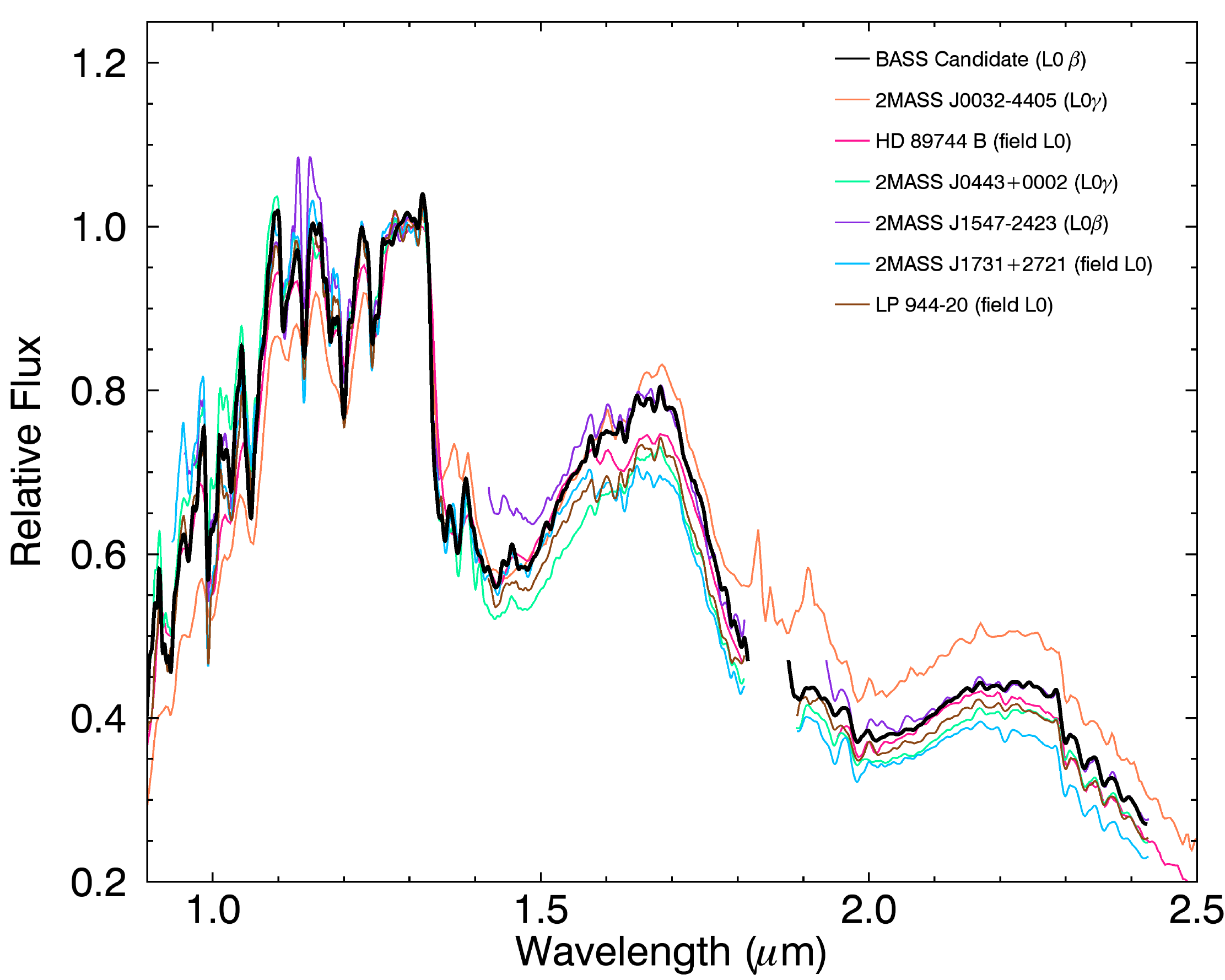}}
	\subfigure[\emph{BASS}~YL002 (L0~$\gamma$)]{\includegraphics[width=0.495\textwidth]{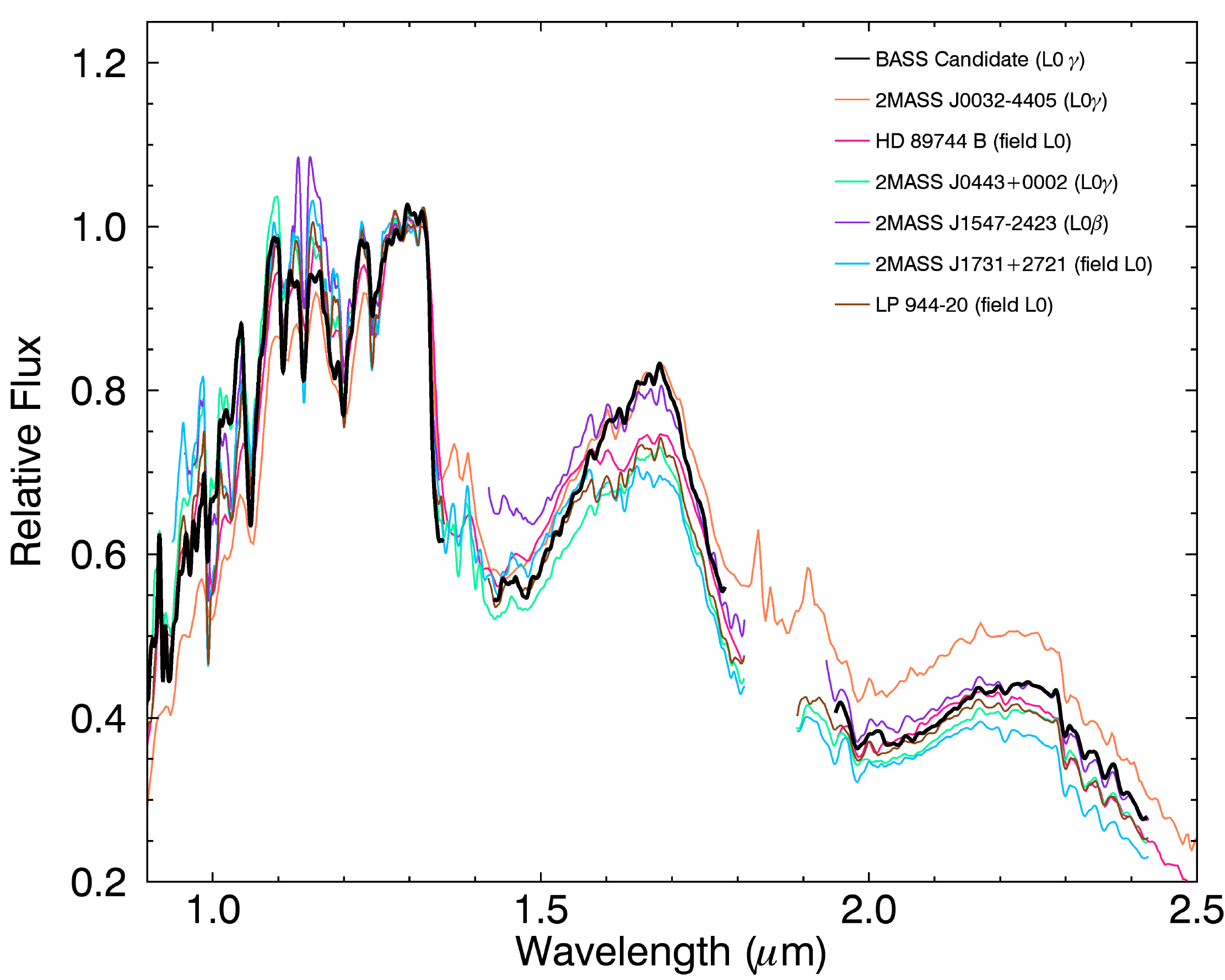}}
	\subfigure[\emph{BASS}~YL003 (L1~$\gamma$)]{\includegraphics[width=0.495\textwidth]{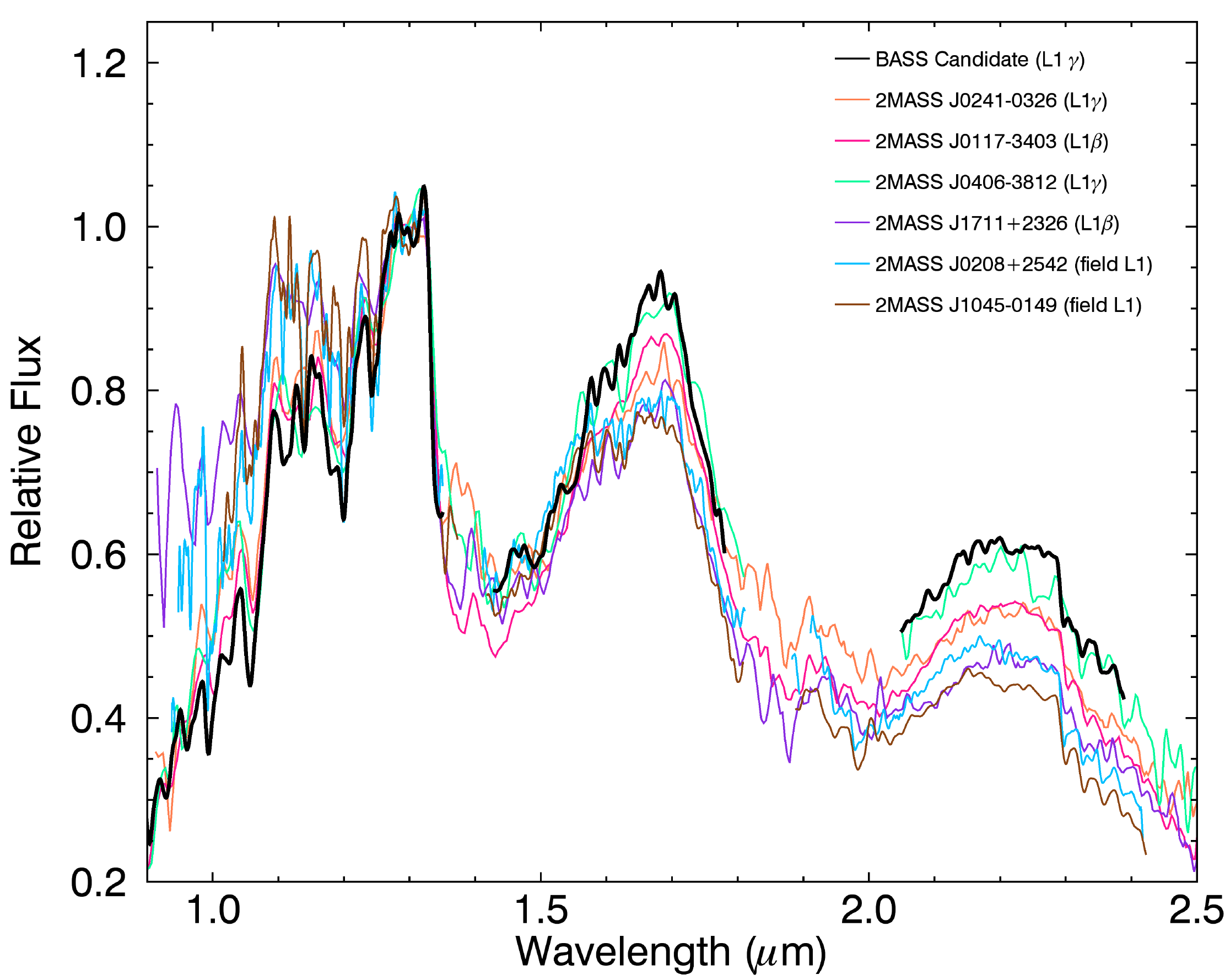}}
	\subfigure[\emph{BASS}~YL004 (L1~$\gamma$)]{\includegraphics[width=0.495\textwidth]{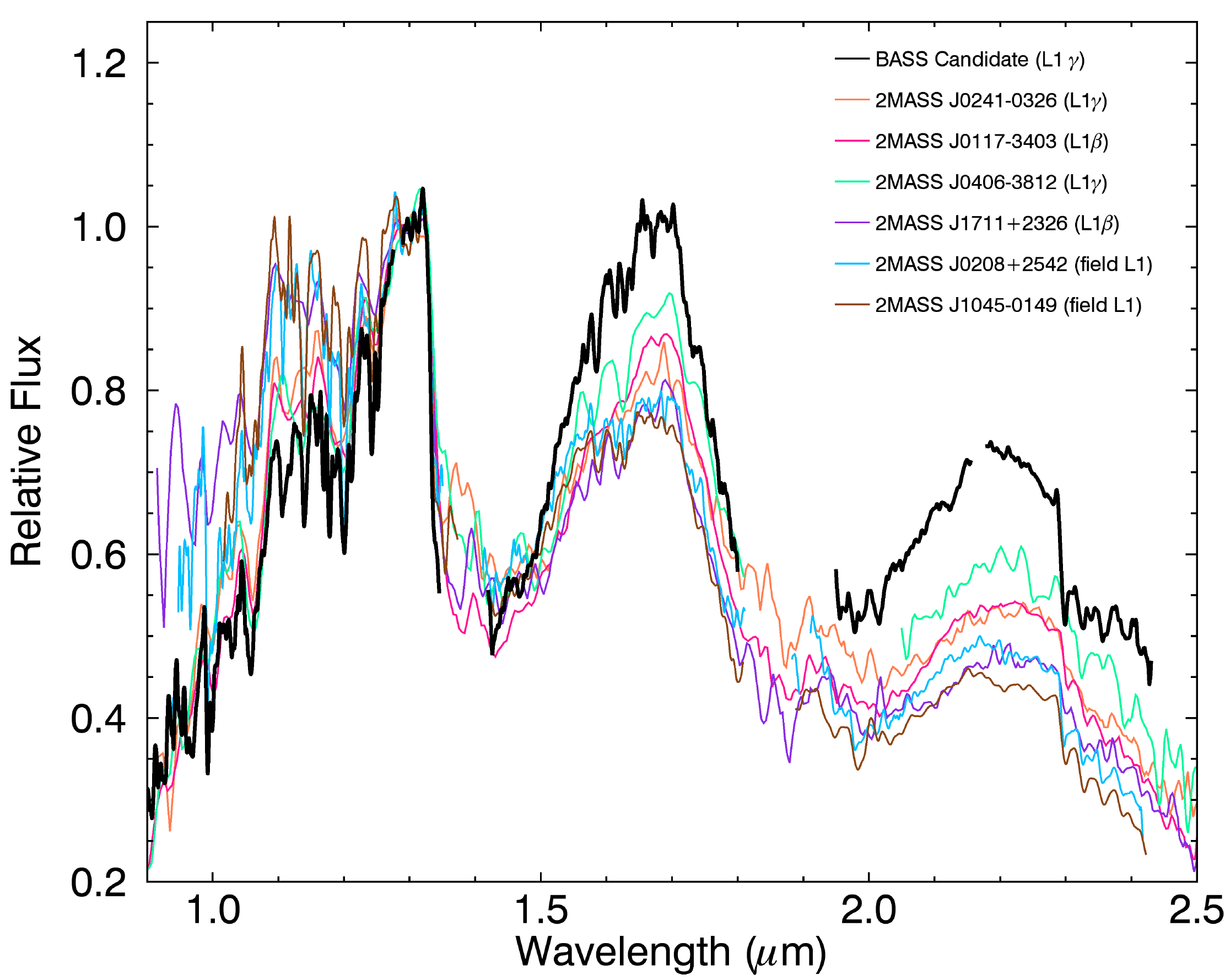}}
	\subfigure[\emph{BASS}~YL005 (L1~$\gamma$)]{\includegraphics[width=0.495\textwidth]{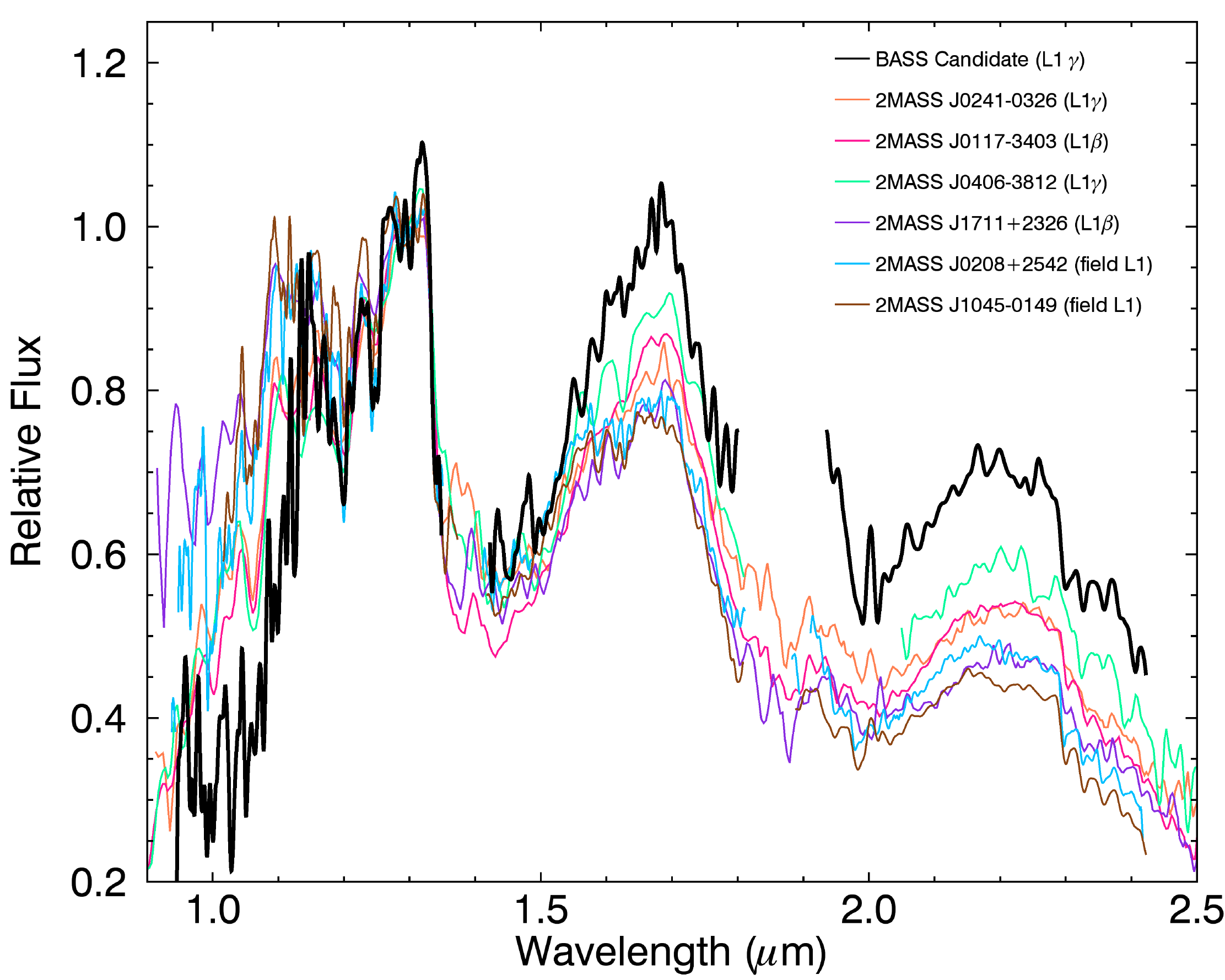}}
	\subfigure[\emph{BASS}~YL006 (L1~$\gamma$)]{\includegraphics[width=0.495\textwidth]{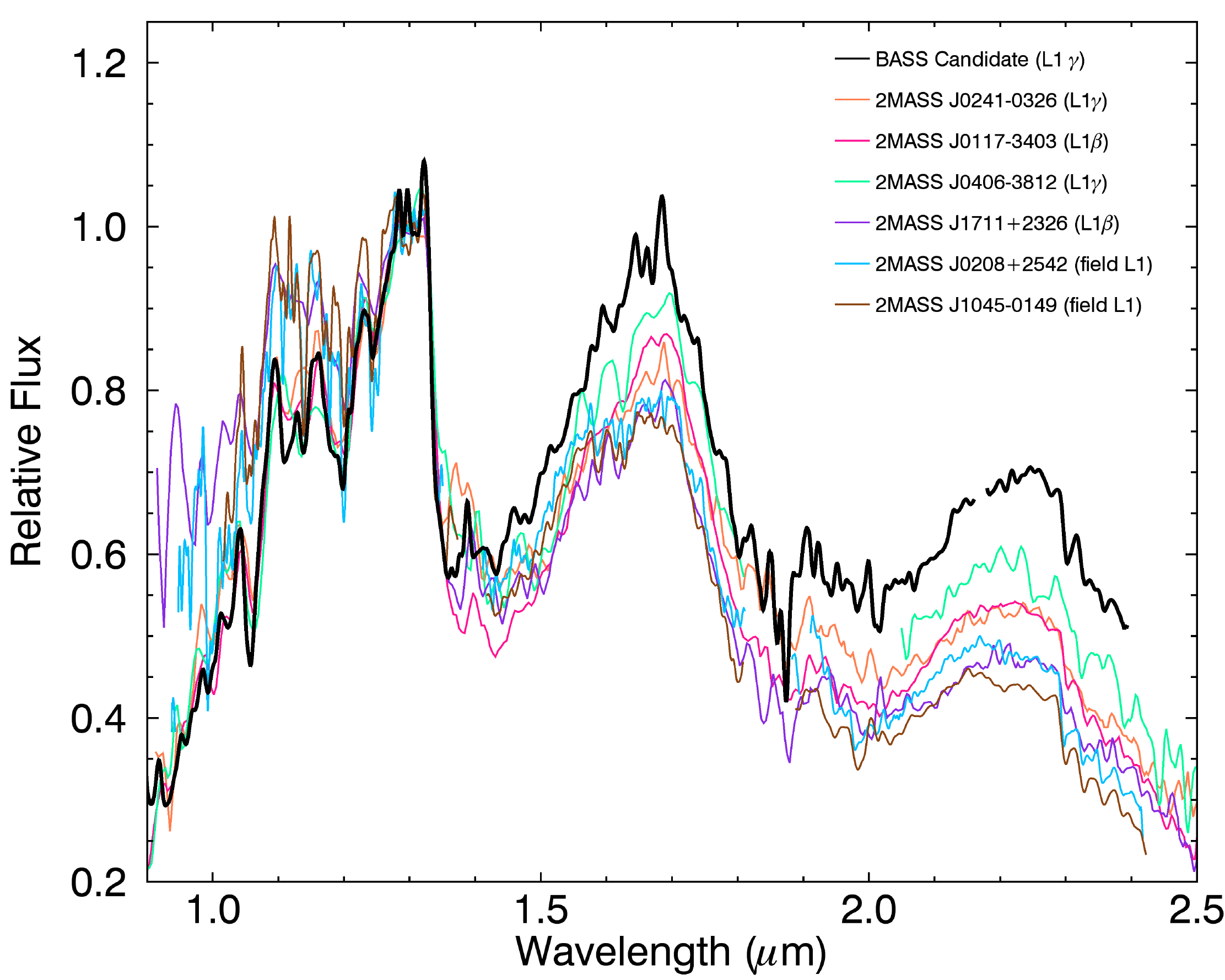}}
	\caption{NIR relative spectral distributions of new young BDs discovered in the \emph{BASS} survey (thick black lines), compared to various field and young BDs of the same spectral types (colored lines). All spectra are normalized at their median value in the 1.27--1.33 $\mu$m range. It can be seen that the new \emph{BASS} discoveries display hallmark signs of low-gravity, such as a redder slope and a triangular continuum shape in the $H$-band (1.45--1.8 $\mu$m).}
	\label{fig:C1}
\end{figure}

%Figure : Candidates, batch 2
\begin{figure}
	\centering
	\subfigure[\emph{BASS}~YL007 (L2~$\beta$)]{\includegraphics[width=0.495\textwidth]{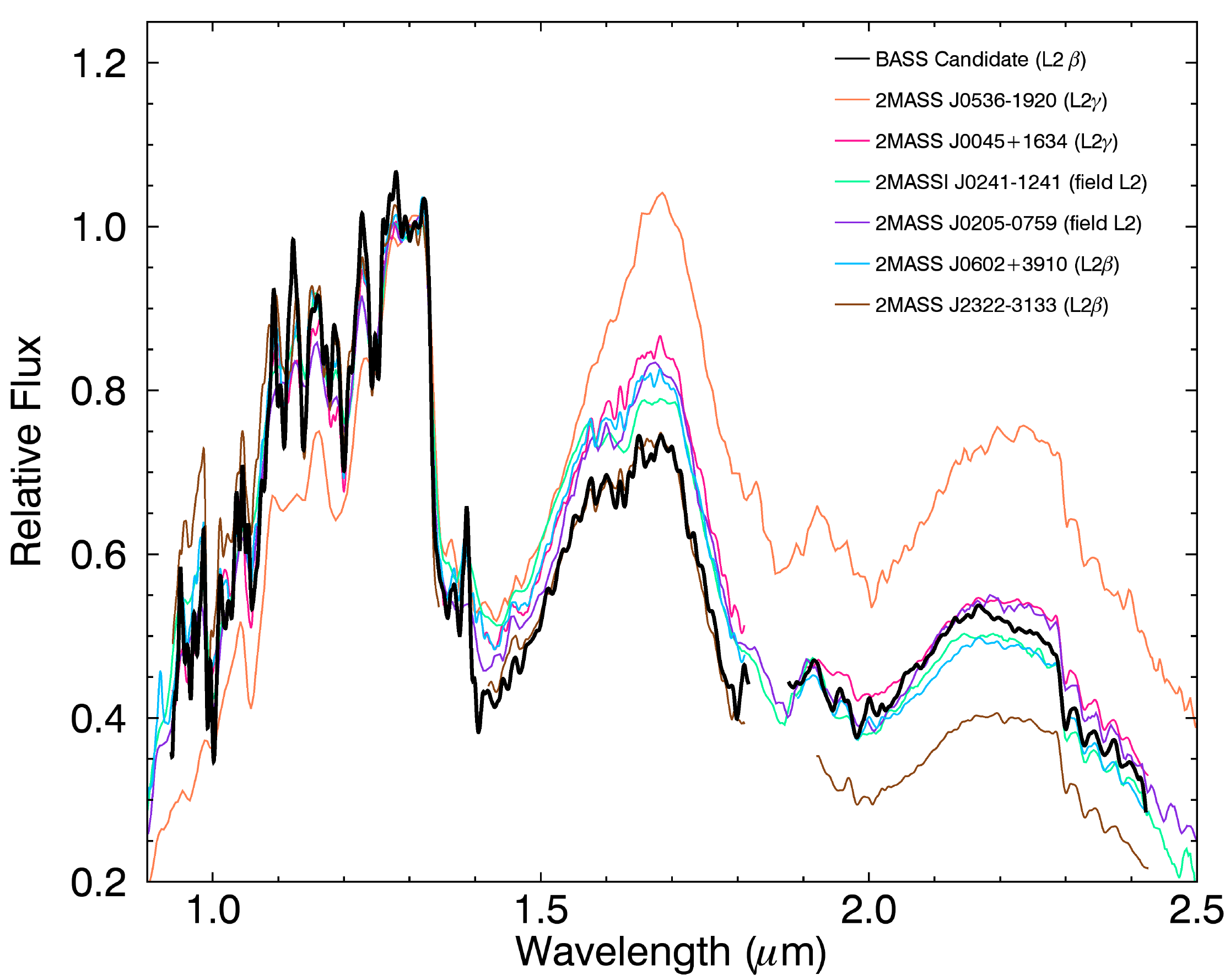}}
	\subfigure[\emph{BASS}~YL008 (L2~$\gamma$)]{\includegraphics[width=0.495\textwidth]{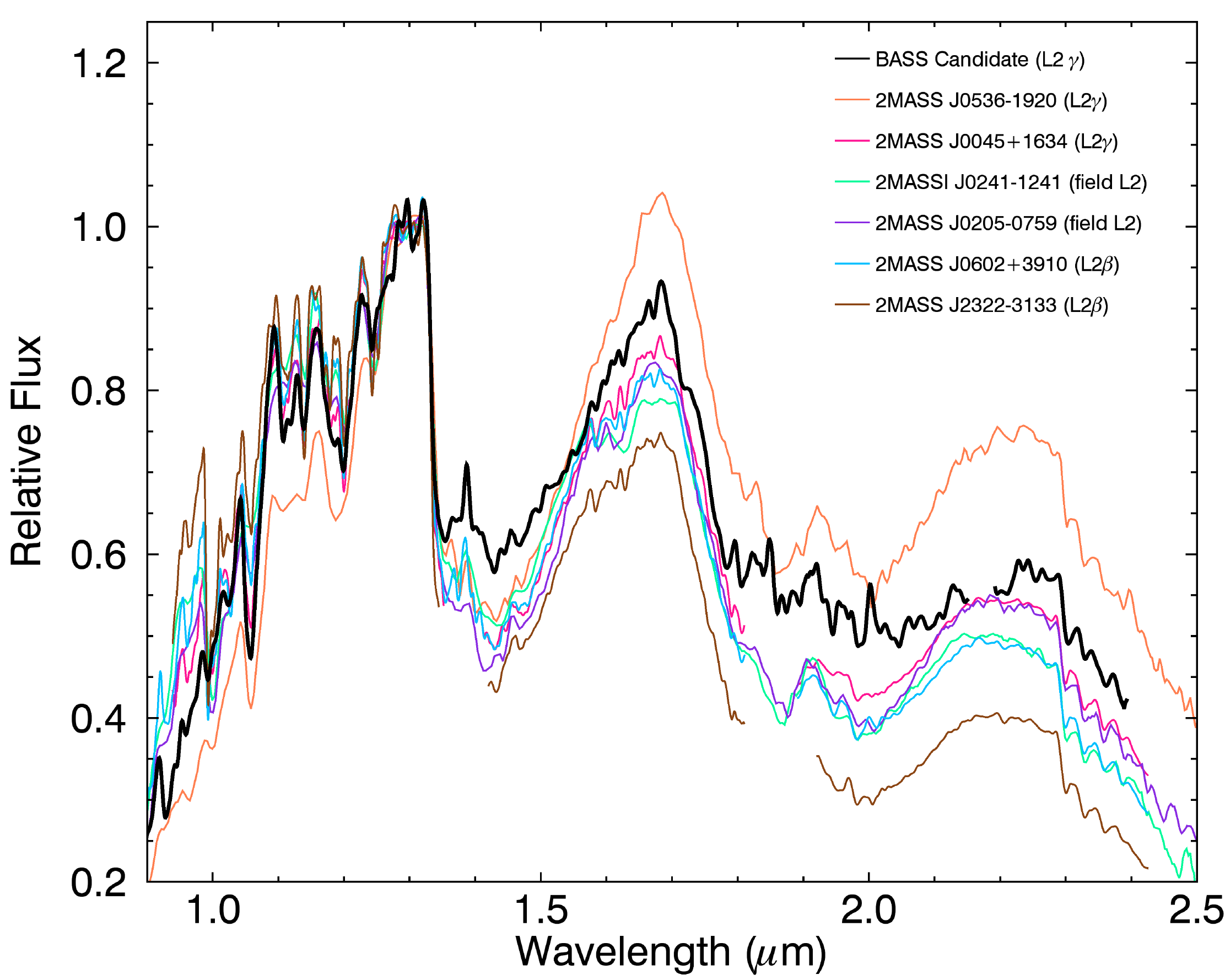}}
	\subfigure[\emph{BASS}~YL009 (L2~$\gamma$)]{\includegraphics[width=0.495\textwidth]{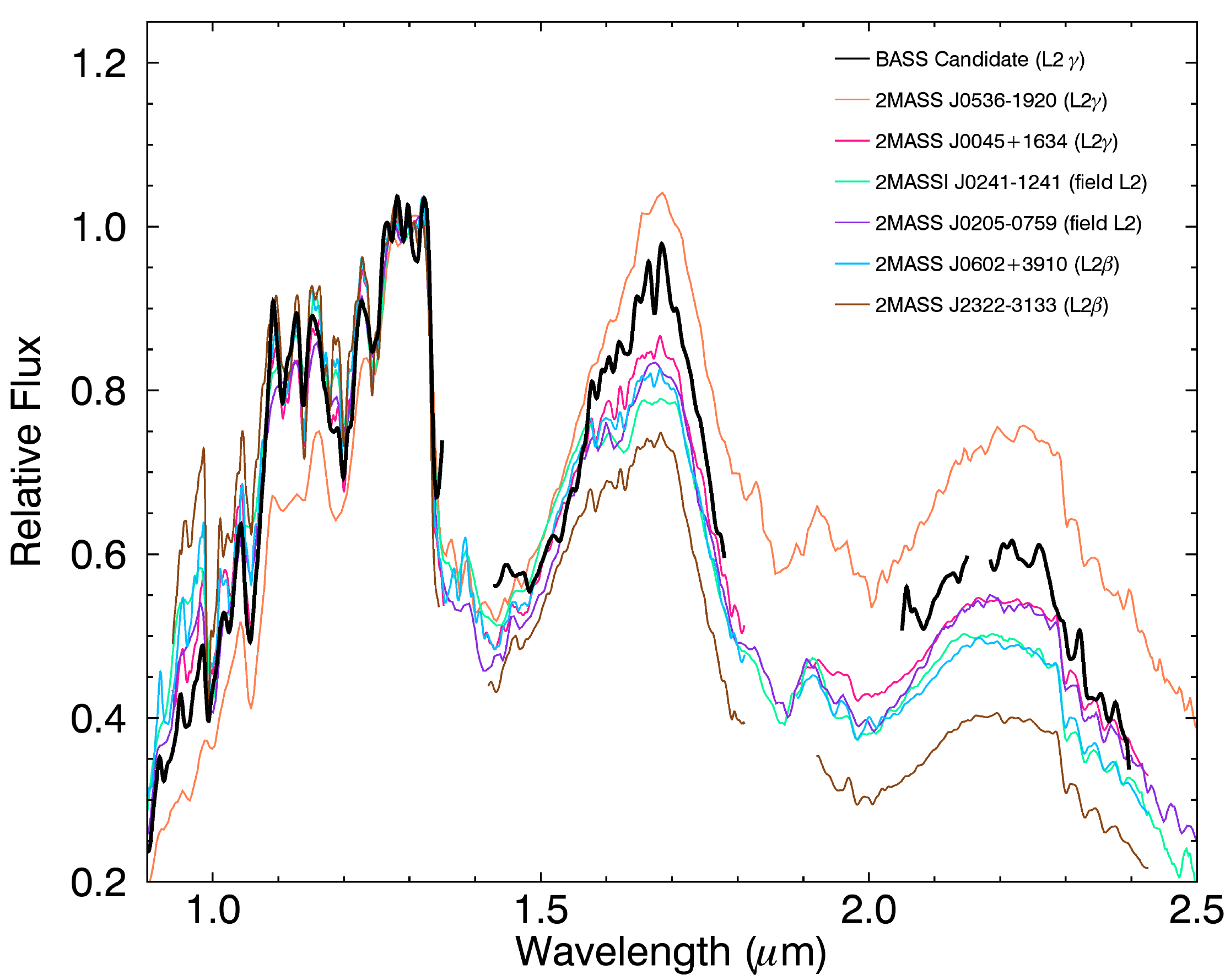}}
	\subfigure[\emph{BASS}~YL010 (L3~$\gamma$)]{\includegraphics[width=0.495\textwidth]{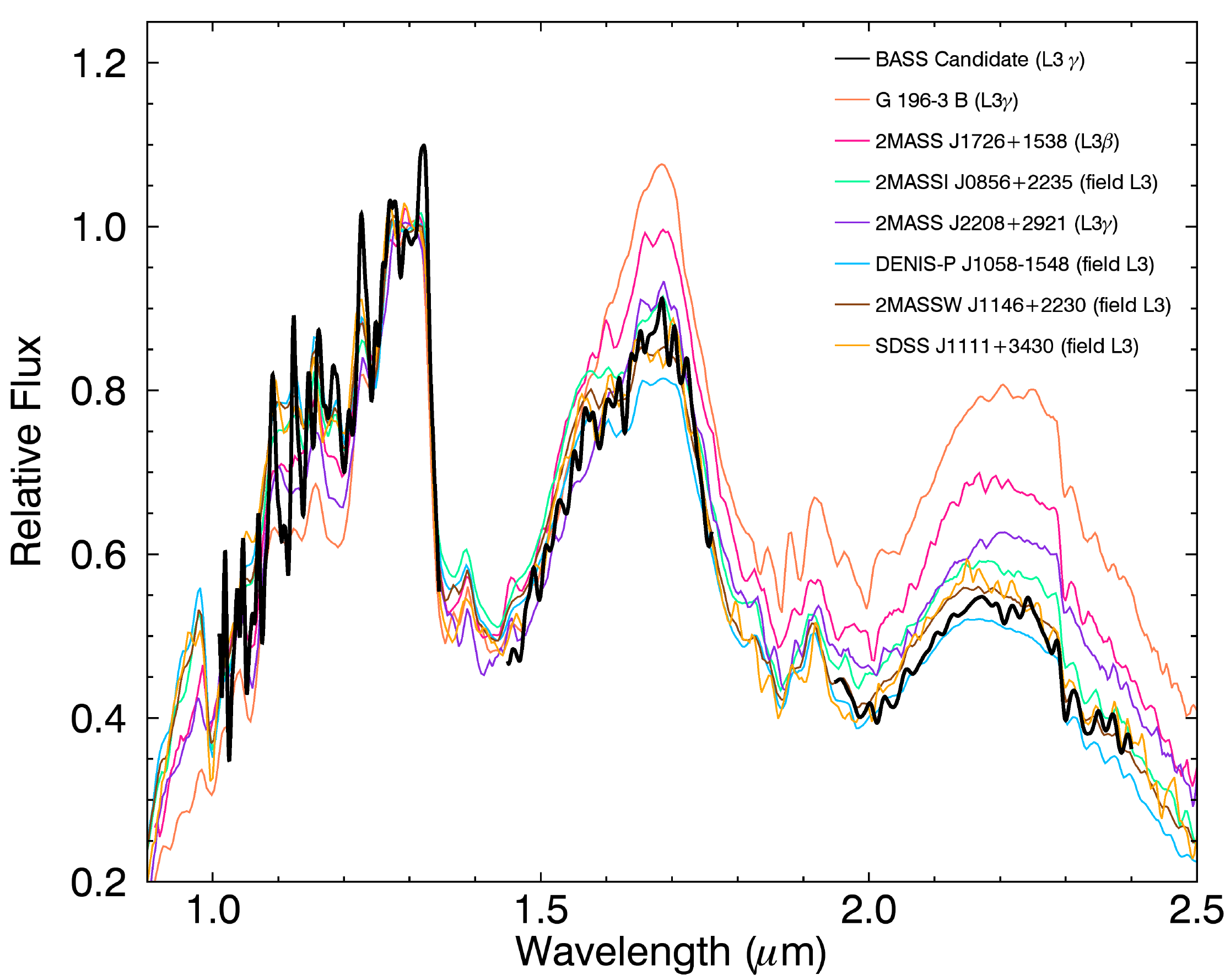}}
	\subfigure[\emph{BASS}~YL011 --- \href{http://simbad.cfa.harvard.edu/simbad/sim-id?Ident=\%408845978&Name=SIMP\%20J215434.5-105530.8}{SIMP~J2154--1055} (L4~$\beta$)]{\includegraphics[width=0.495\textwidth]{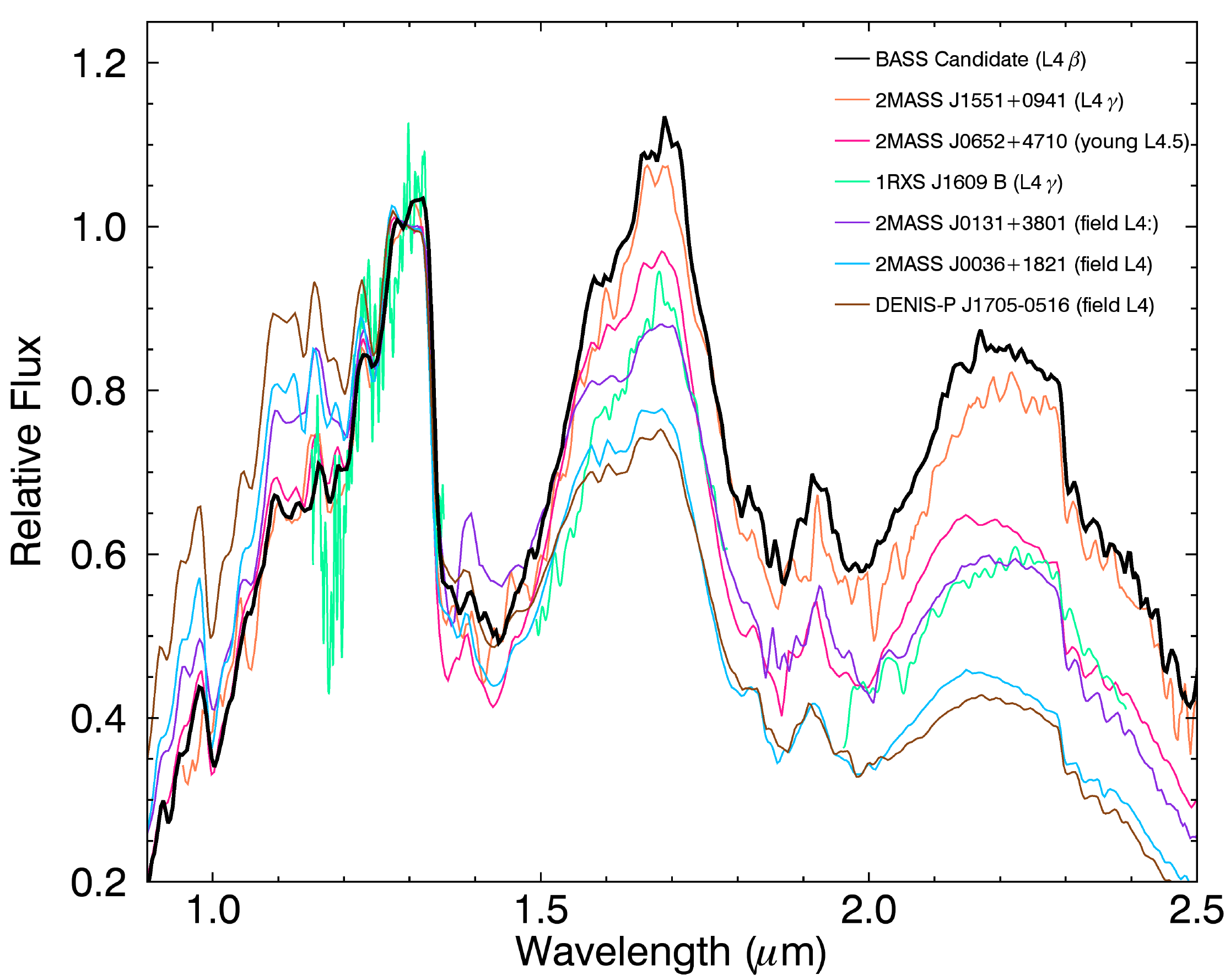}}
	\subfigure[\emph{BASS}~YL012 (L4~$\beta$)]{\includegraphics[width=0.495\textwidth]{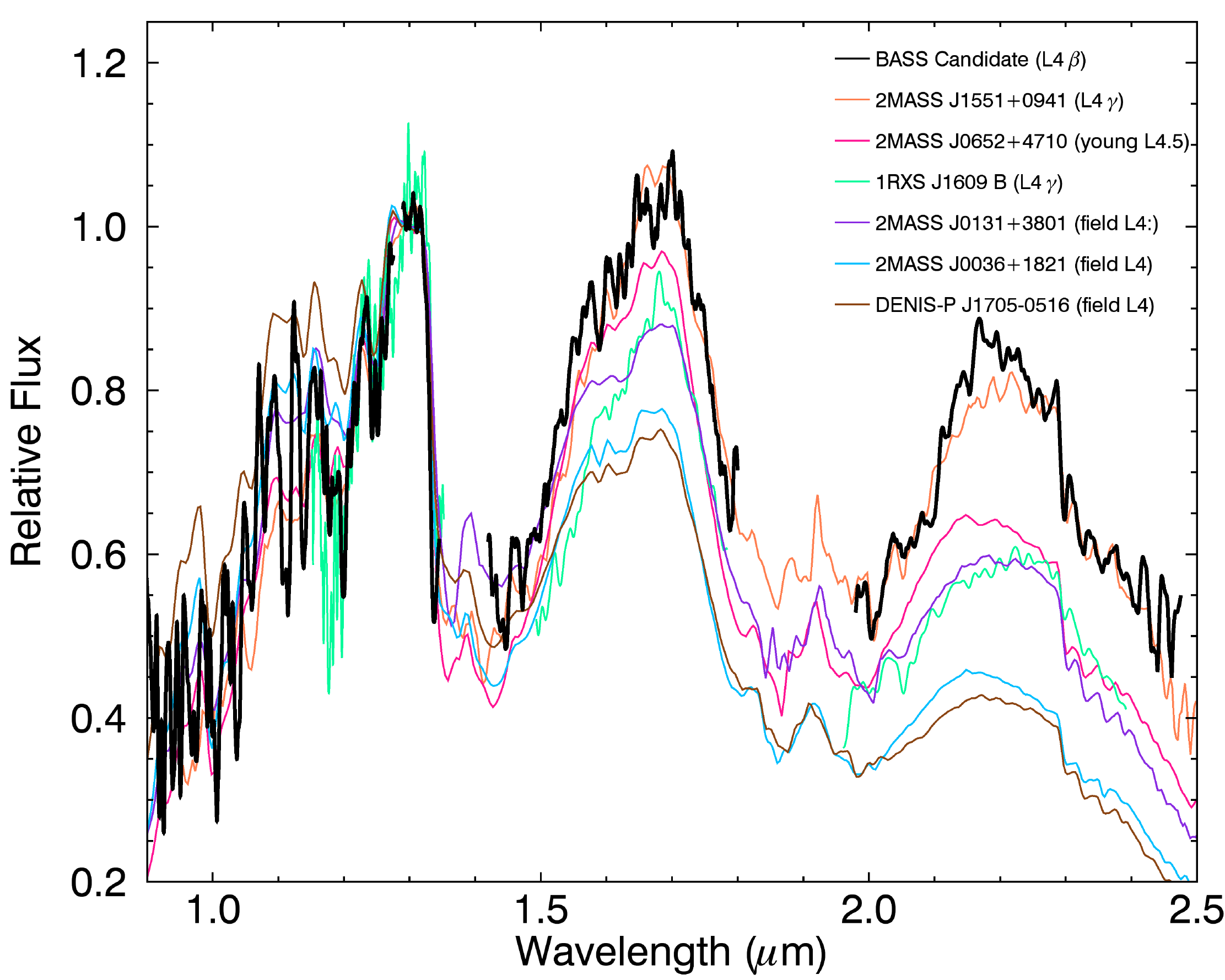}}
\caption{NIR relative spectral distributions of new young BDs discovered in the \emph{BASS} survey, compared to various field and young BDs of the same spectral type (see \hyperref[fig:C1]{Figure~\ref*{fig:C1}} for color codes). The young L4~$\beta$ \href{http://simbad.cfa.harvard.edu/simbad/sim-id?Ident=\%408845978&Name=SIMP\%20J215434.5-105530.8}{SIMP~J2154--1055} (panel e) has been independently discovered \citep{2014ApJ...792L..17G} in the \emph{Survey Infrarouge de Mouvement Propre} survey (\emph{SIMP}; \citealp{2009AIPC.1094..493A}; \href{http://www.astro.umontreal.ca/\textasciitilde gagne/simp.php}{J.~Robert et al., in preparation}).}
	\label{fig:C2}
\end{figure}

\section{RESULTS FROM THE \emph{BASS} SURVEY}\label{sec:res}

We present in this section preliminary results from the \emph{BASS} survey: this includes a number of new young BDs (\hyperref[sec:newobj]{Section~\ref*{sec:newobj}}), a new planetary companion to a low-mass star (\hyperref[sec:planet]{Section~\ref*{sec:planet}}) and tentative indications of mass segregation in some YMGs (\hyperref[sec:mseg]{Section~\ref*{sec:mseg}}).

\subsection{New Low-Mass Stars and Brown Dwarfs with Signs of Low-Gravity}\label{sec:newobj}

\begin{figure}
	\centering
	\subfigure[FeH$_Z$ ($R \geq 75$)]{\includegraphics[width=0.495\textwidth]{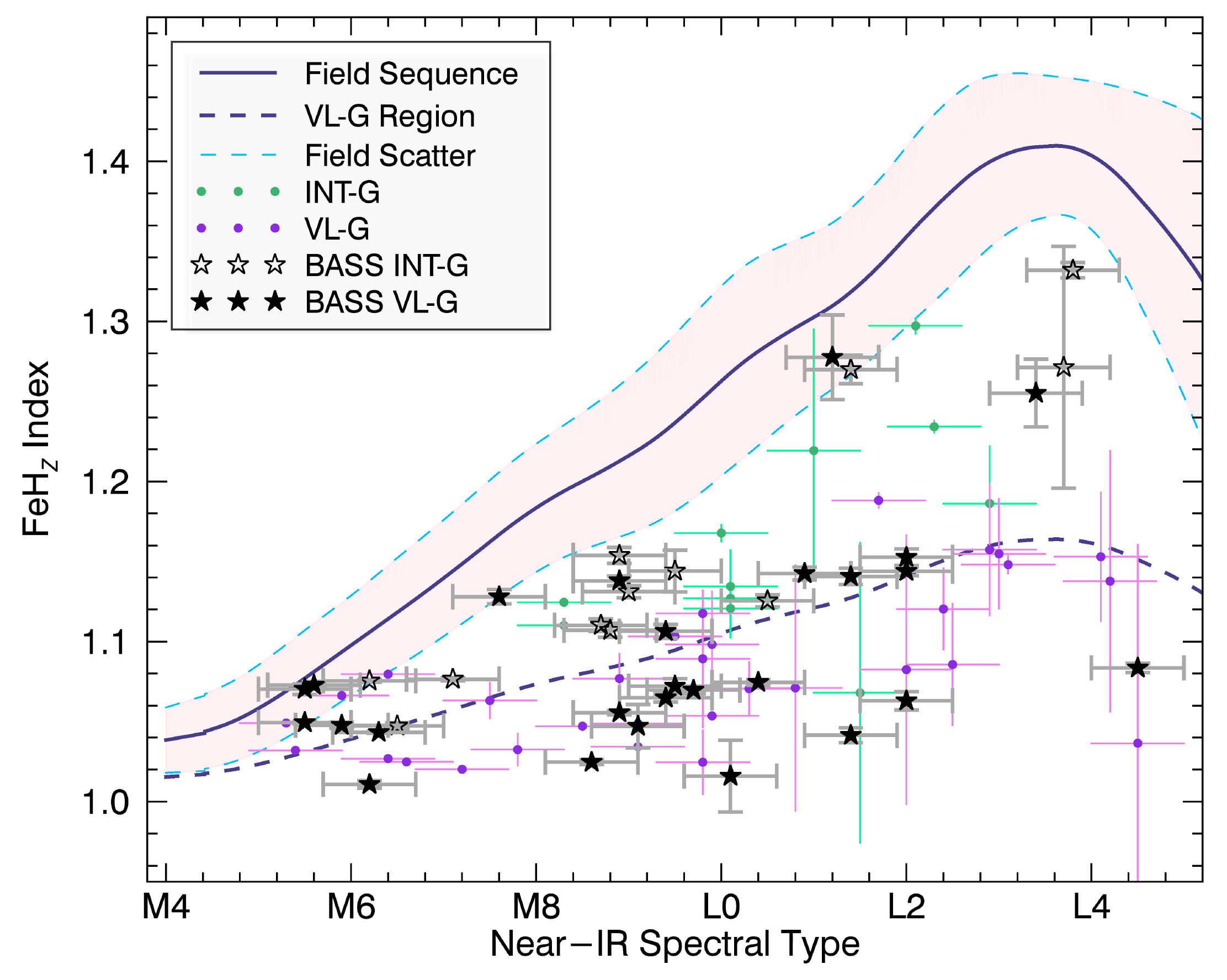}}
	\subfigure[$H$-cont ($R \geq 75$)]{\includegraphics[width=0.495\textwidth]{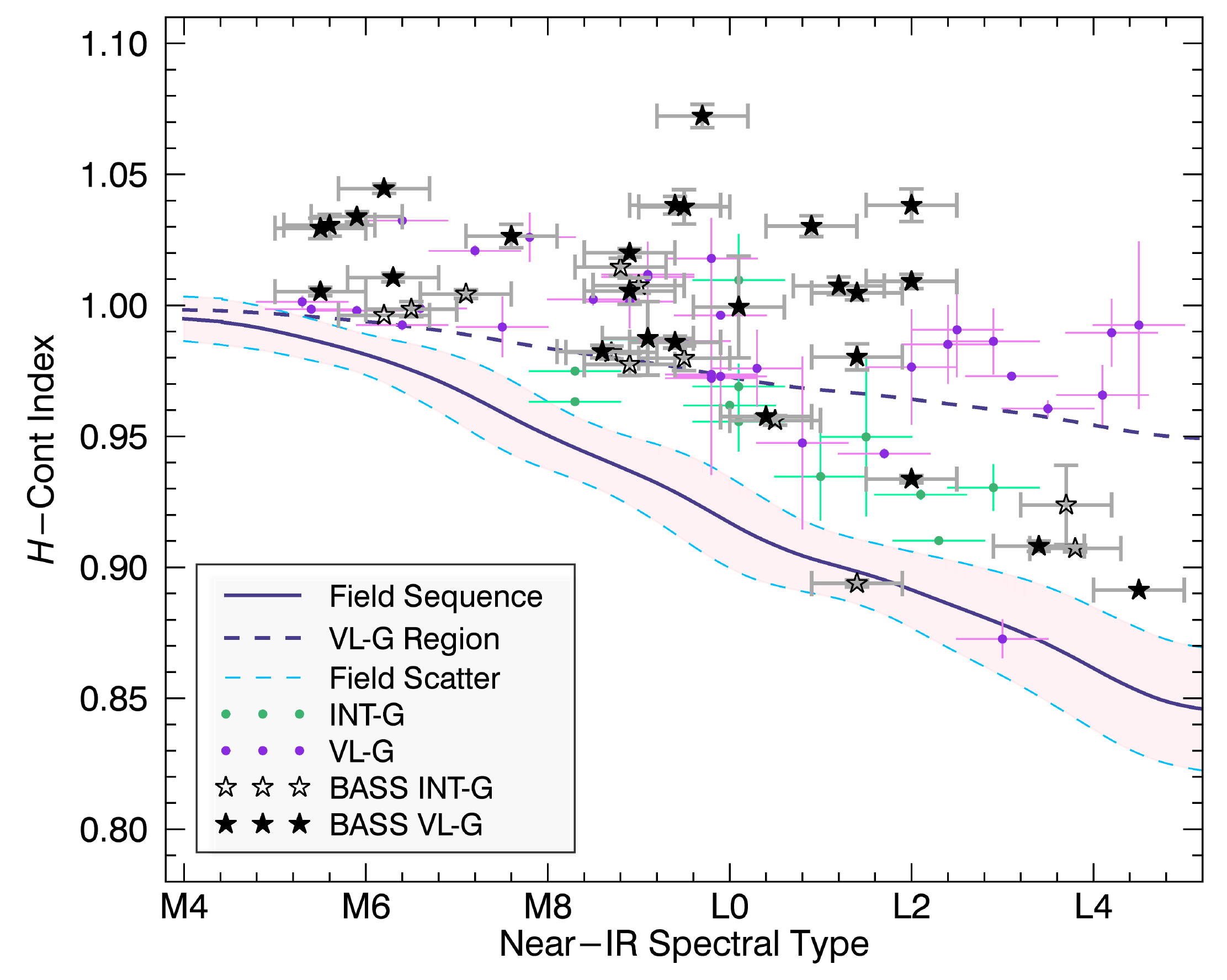}}
	\subfigure[VO$_Z$ ($R \geq 75$)]{\includegraphics[width=0.495\textwidth]{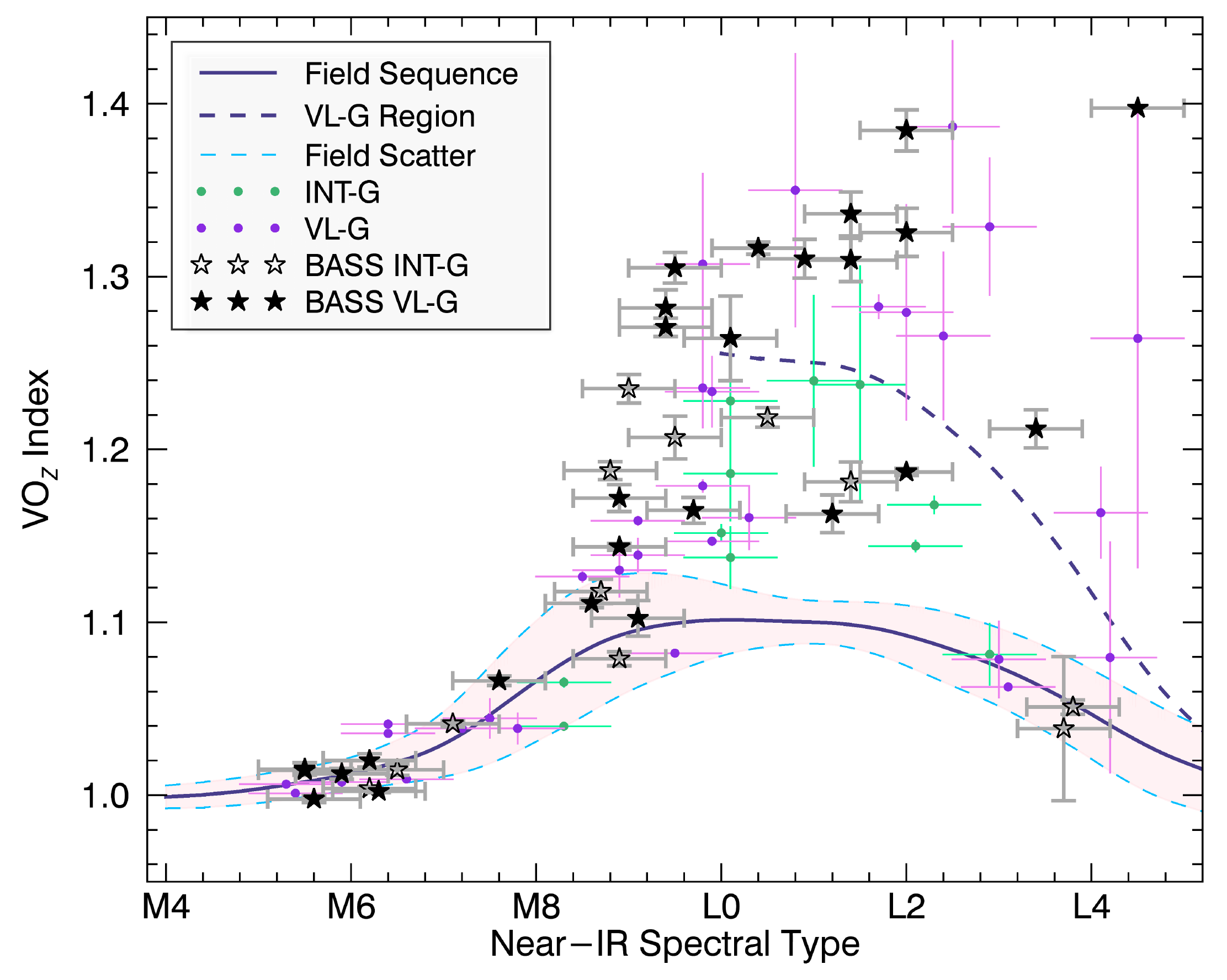}}
	\subfigure[$J - K_S$]{\includegraphics[width=0.495\textwidth]{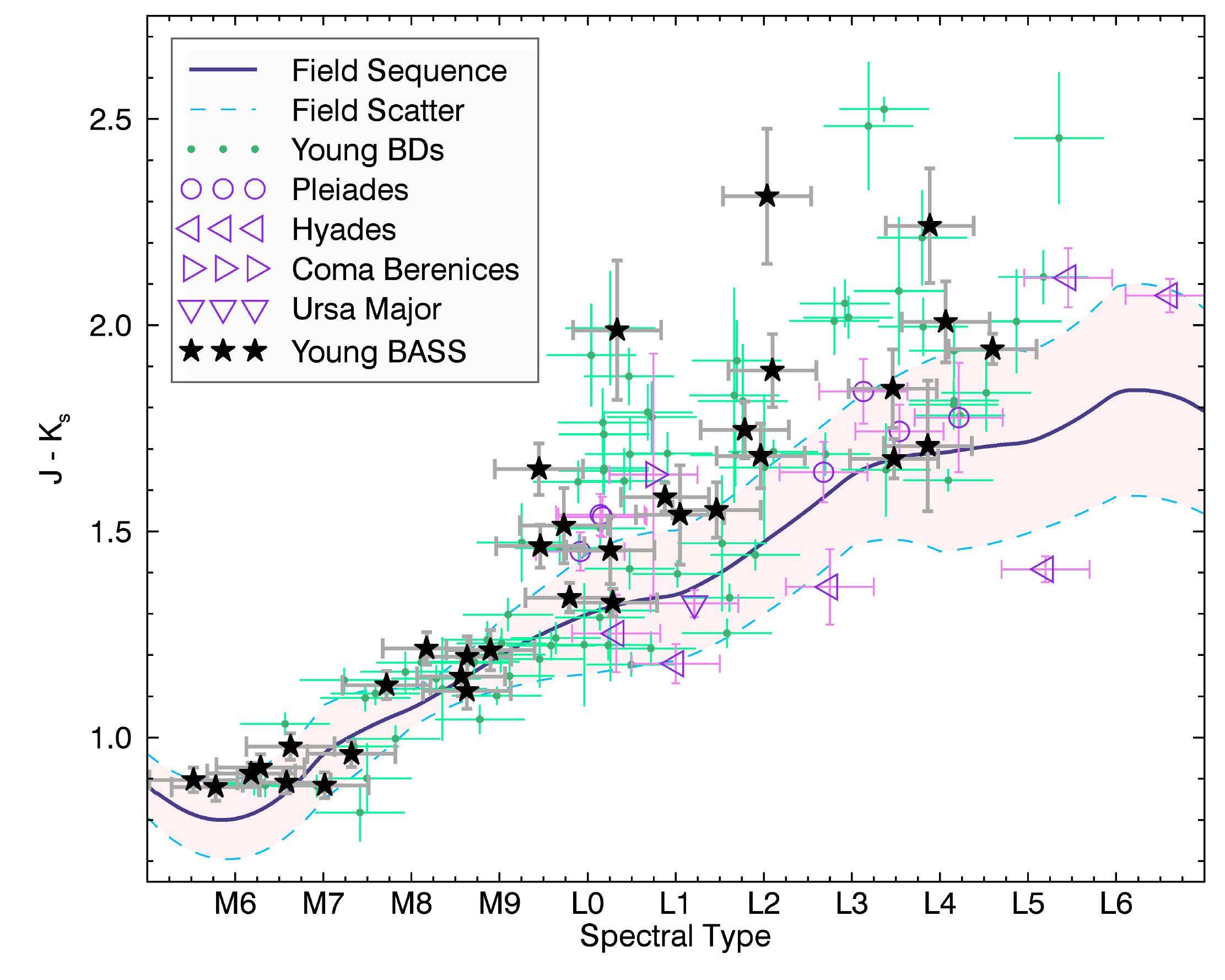}}
	\caption{Panels (a)--(c): Gravity-sensitive NIR spectroscopic indices of new discoveries in the \emph{BASS} survey (open black stars classified as INT-G and filled black stars classified as VL-G; see text), compared to the field sequence (thick blue line) and its scatter (pink region delimited by dashed pale blue lines) as defined in \cite{2013ApJ...772...79A}. The thick, dashed blue line represents the delimitation between the intermediate and very low-gravity regimes and known young BDs are represented with filled dots (green dots for INT-G and purple dots for VL-G). Random offsets with a standard deviation of 0.2 subtypes were added to spectral types for visibility. Panel (d): \emph{2MASS} $J-K_S$ colors as a function of spectral type of new young BD discoveries from the \emph{BASS} survey (filled black stars), compared to the field sequence (blue line) and its scatter (pink region delimited by dashed pale blue lines). The field sequence was built from the DwarfArchive. Known young field BDs are displayed as green dots, whereas members of slightly older associations are displayed as open purple symbols; circles for the Pleiades, downside triangles for Ursa Major ($\sim$400 Myr), right triangles for Coma Berenices ($\sim$500 Myr) and left triangles for the Hyades ($\sim$625 Myr). Young L dwarfs generally display redder NIR colors compared to their older equivalents due to an enhanced atmospheric dust thickness.}
	\label{fig:I1}
\end{figure}

\begin{figure}
	\centering
	\subfigure[KI$_J$ ($R \geq 75$)]{\includegraphics[width=0.495\textwidth]{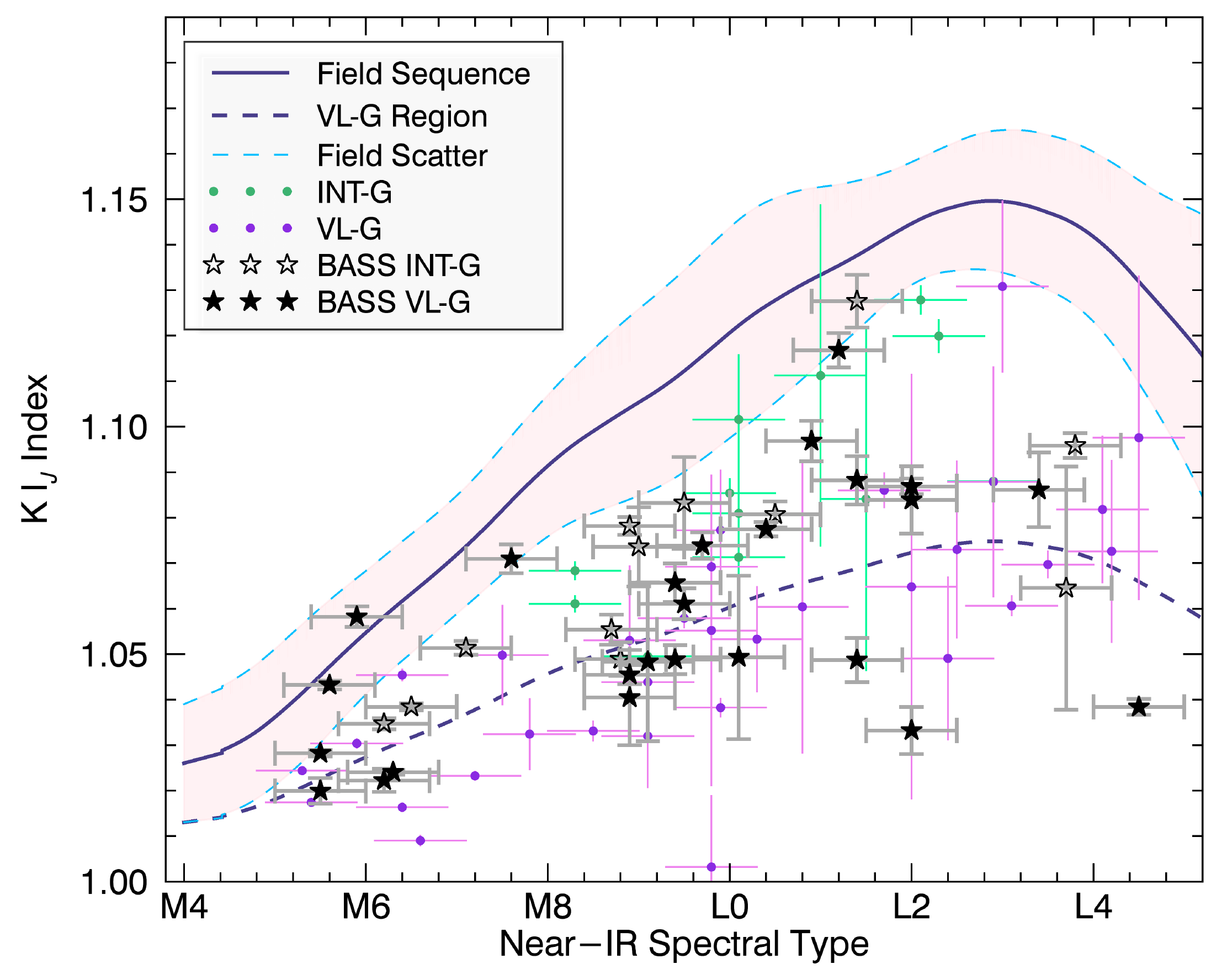}}
	\subfigure[FeH$_J$ ($R \geq 750$)]{\includegraphics[width=0.495\textwidth]{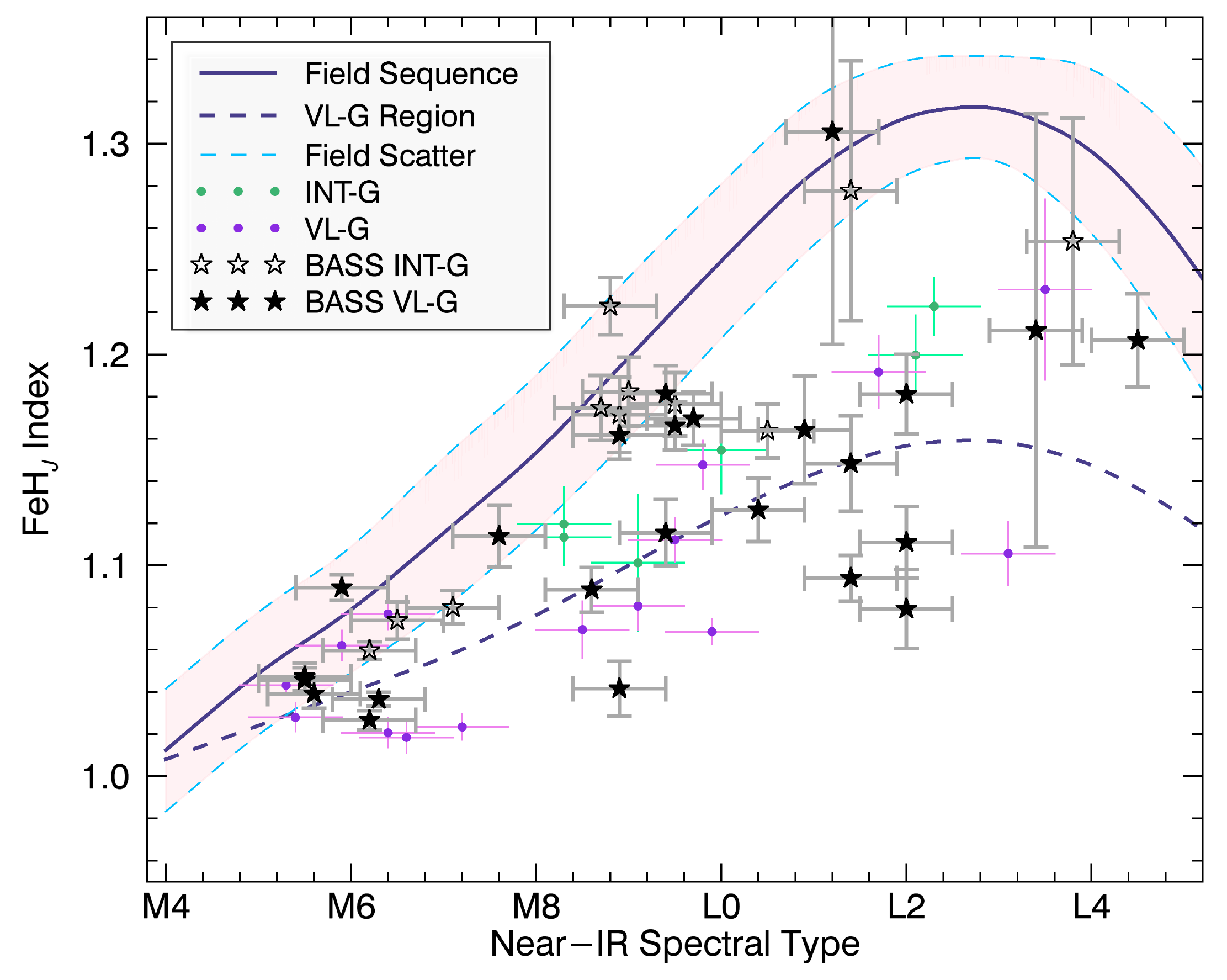}}
	\subfigure[K~I 1.169 $\mu$m equivalent width ($R \geq 750$)]{\includegraphics[width=0.495\textwidth]{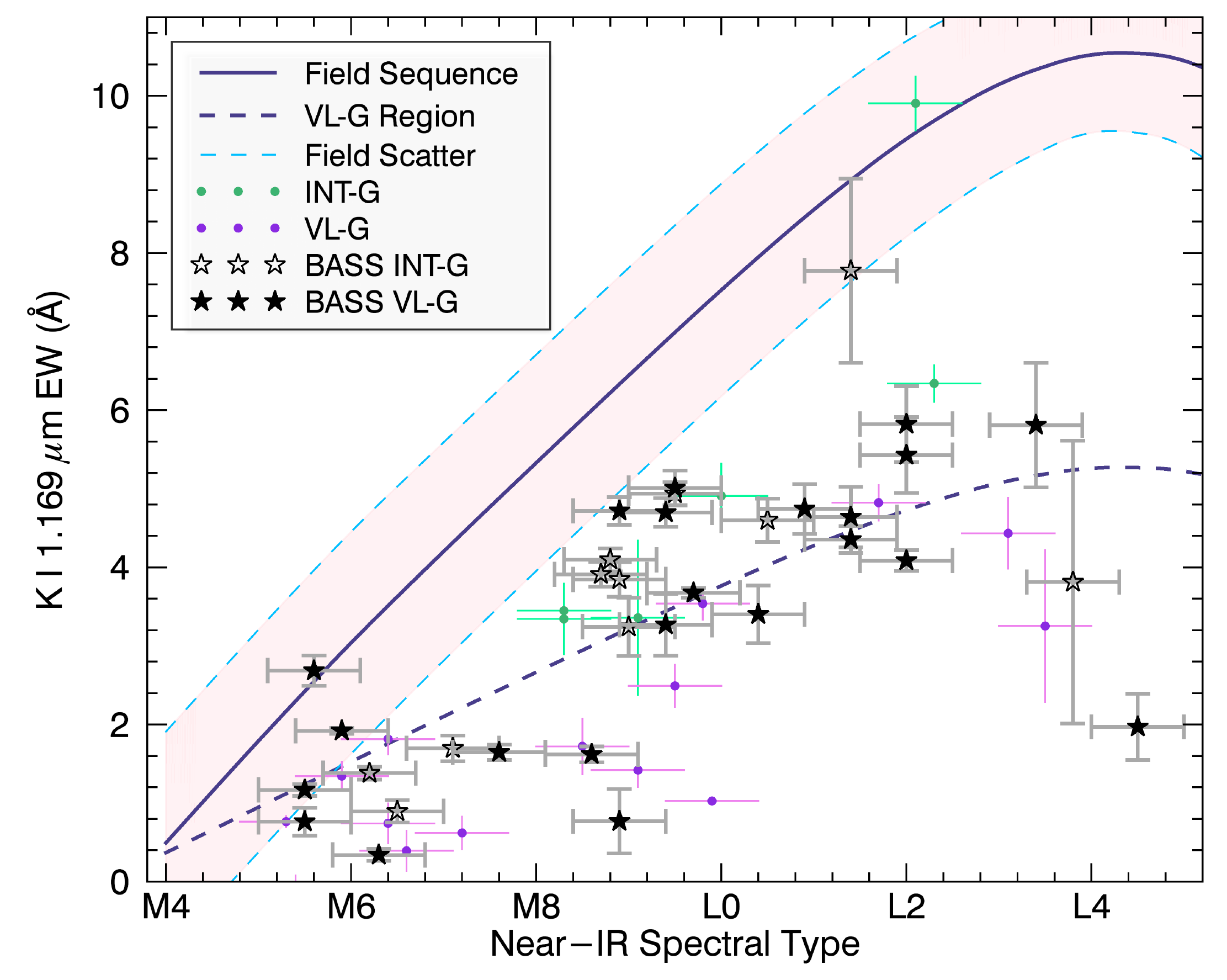}}
	\subfigure[K~I 1.244 $\mu$m equivalent width ($R \geq 750$)]{\includegraphics[width=0.495\textwidth]{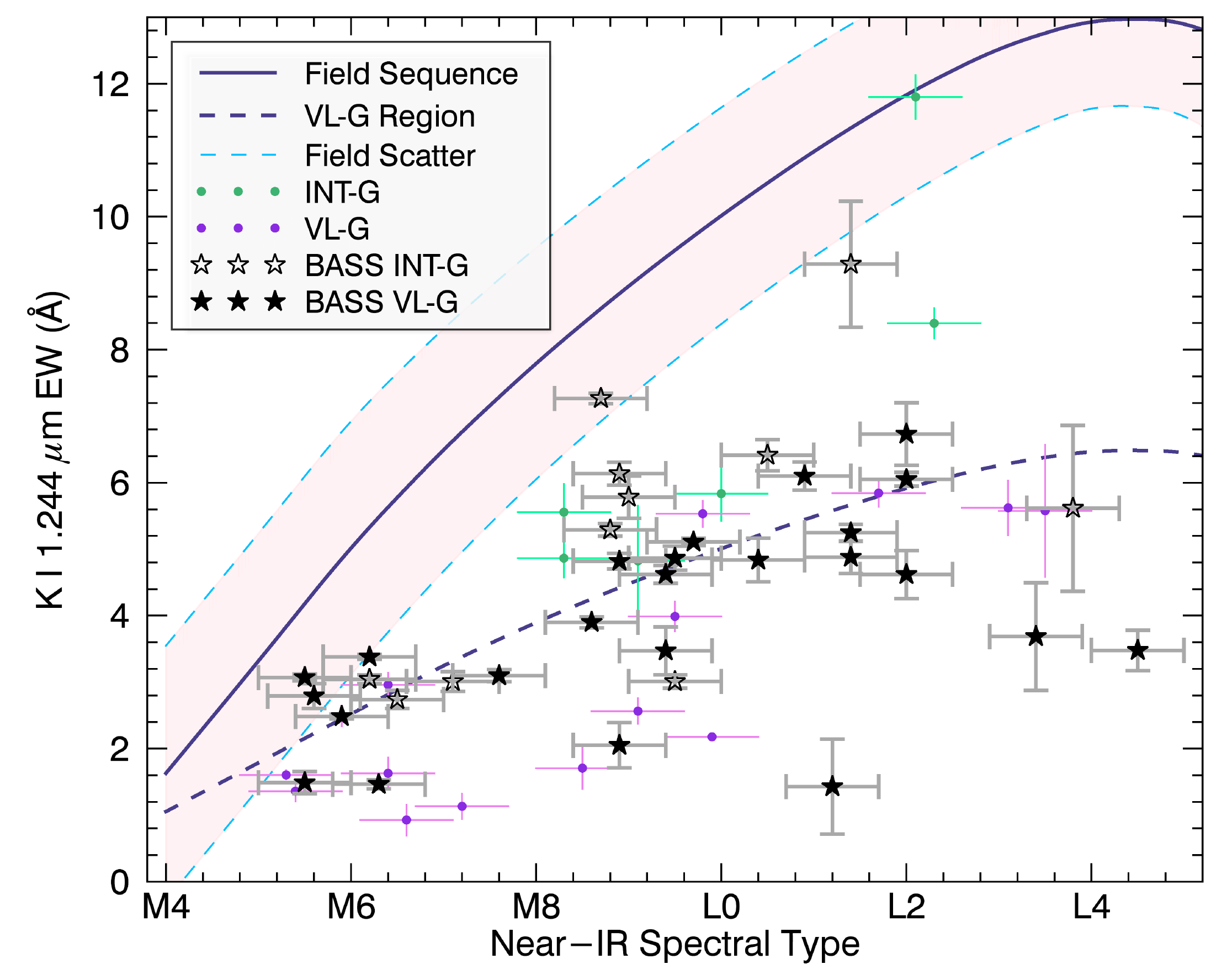}}
	\subfigure[K~I 1.253 $\mu$m equivalent width ($R \geq 750$)]{\includegraphics[width=0.495\textwidth]{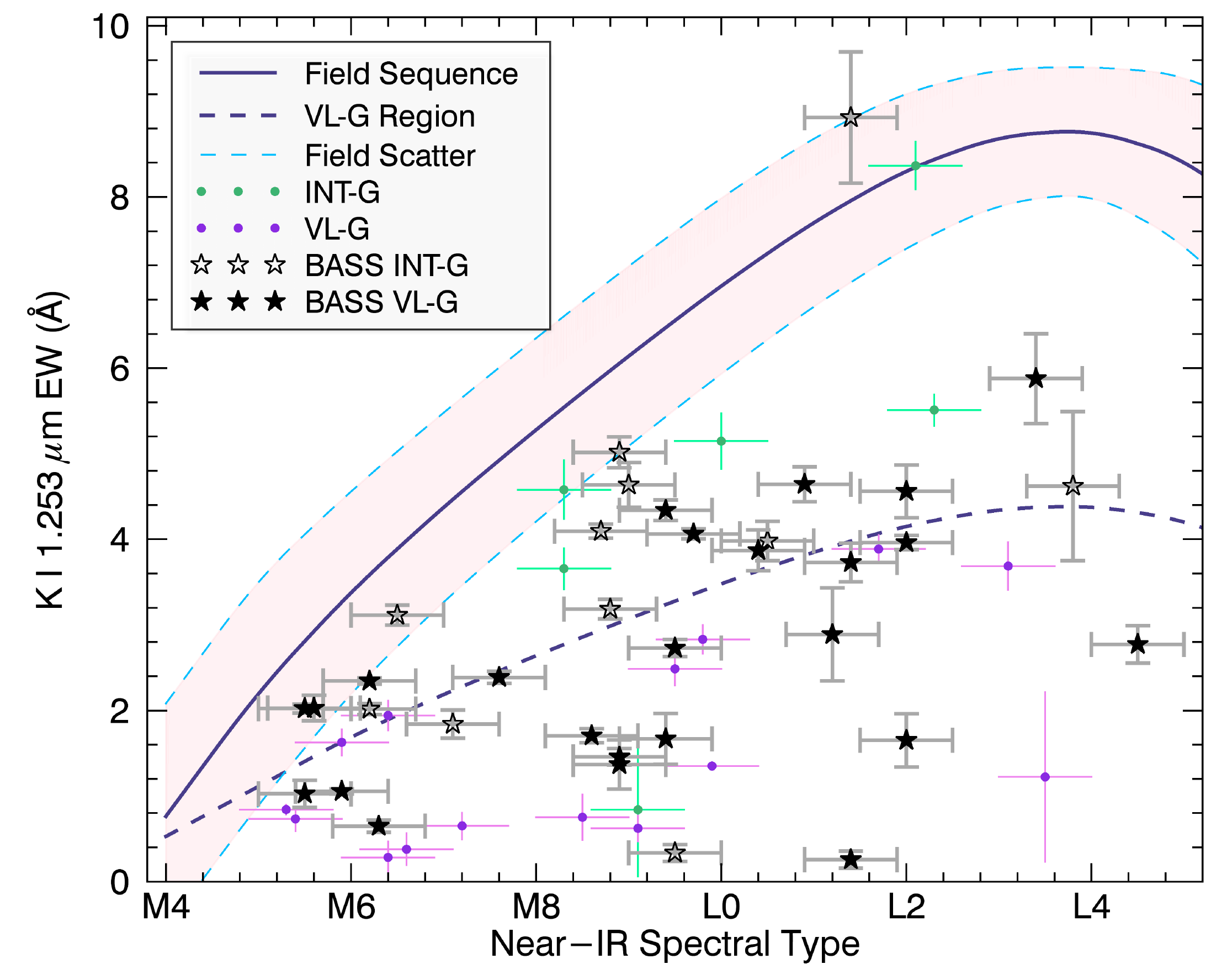}}
	\subfigure[Na~I 1.138 $\mu$m equivalent width ($R \geq 750$)]{\includegraphics[width=0.495\textwidth]{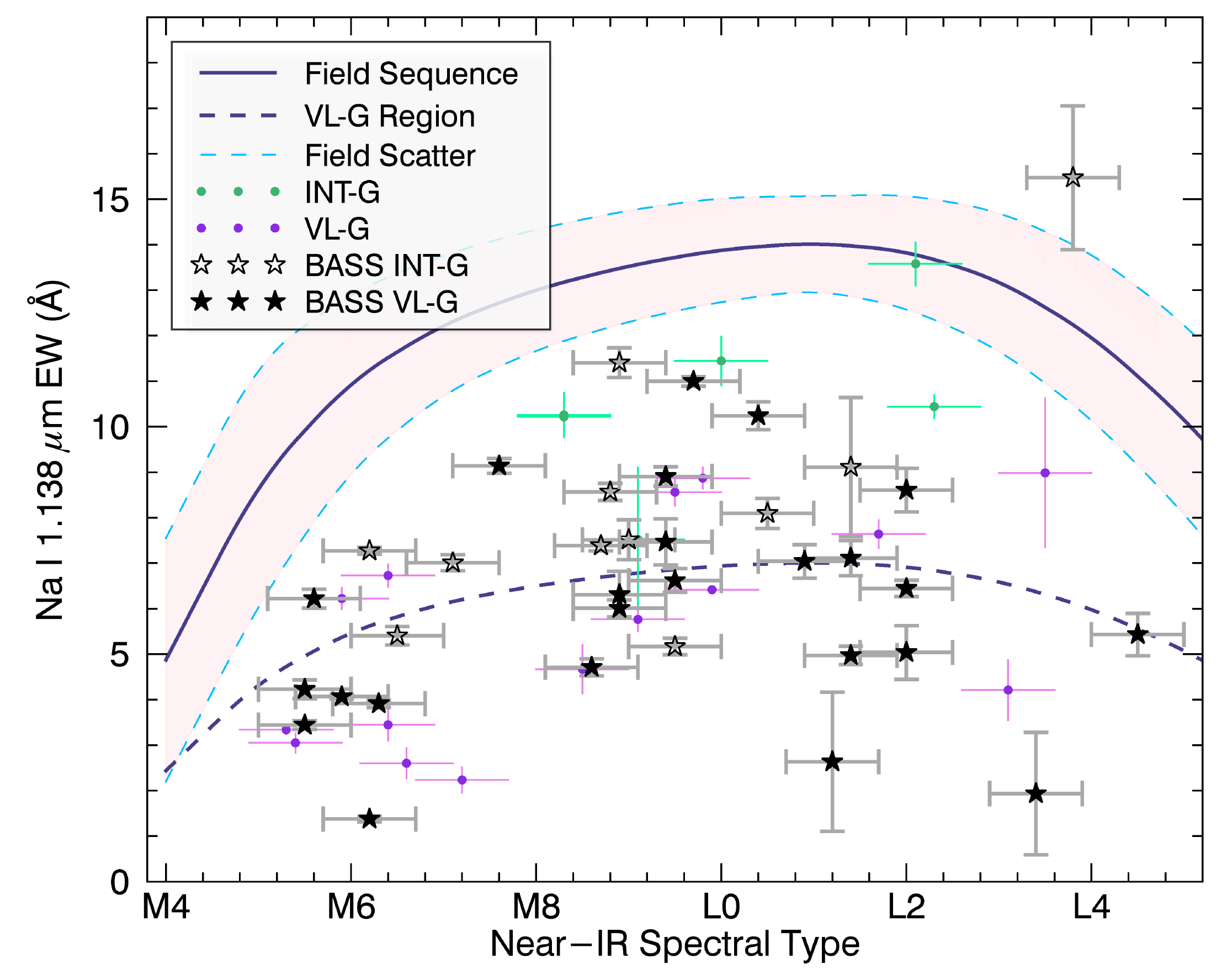}}
	\caption{Gravity-sensitive spectroscopic indices (panels a--b) and equivalent widths (panels c--f) of new young BD discoveries in the \emph{BASS} survey, compared to other young BDs and the field sequence. See \hyperref[fig:I1]{Figure~\ref*{fig:I1}} (panels a--c) for a description of the symbols.}
	\label{fig:I2}
\end{figure}

We identified more than 45 new BDs displaying signs of low gravity (youth), some of which are presented in Figures \ref{fig:C1} and \ref{fig:C2}. We used both a visual and an index-based classification to assess signs of low-gravity. Visual comparison was done using the method of \cite{2009AJ....137.3345C} in the optical and \href{http://www.astro.umontreal.ca/\textasciitilde gagne/NextCruzOct2014.php}{K.~Cruz et al.} (\href{http://www.astro.umontreal.ca/\textasciitilde gagne/NextCruzOct2014.php}{in preparation};  see also \citealp{DisentanglingLDwar:db}) in the NIR: the $J$, $H$ and $K$ bands are normalized individually to ensure that the classification is not affected by the large spread in NIR colors of BDs. We used the NIR classification scheme of \cite{2013ApJ...772...79A} to confirm the visual classifications, and found that both methods agree very well. In Figures \ref{fig:I1} and \ref{fig:I2}, we compare the gravity-sensitive indices of young \emph{BASS} candidates to sequences defined by \cite{2013ApJ...772...79A}. These Figures demonstrate how the \emph{BASS} survey represents a significant contribution to the set of currently known young BDs. More importantly, all these new young BDs are potentially age-calibrated members of nearby moving groups, which will be of crucial importance in understanding how their properties correlate with age. \cite{2013ApJ...772...79A} note that their respective gravity classes FLD-G, INT-G and VL-G seem to correspond to respective ages of $>$ 200 Myr, 50--200 Myr and 10--30 Myr, using 25 young BDs with age constraints. \\

\begin{figure}
	\centering
	\subfigure[Na-a index ($R \geq 750$)]{\includegraphics[width=0.495\textwidth]{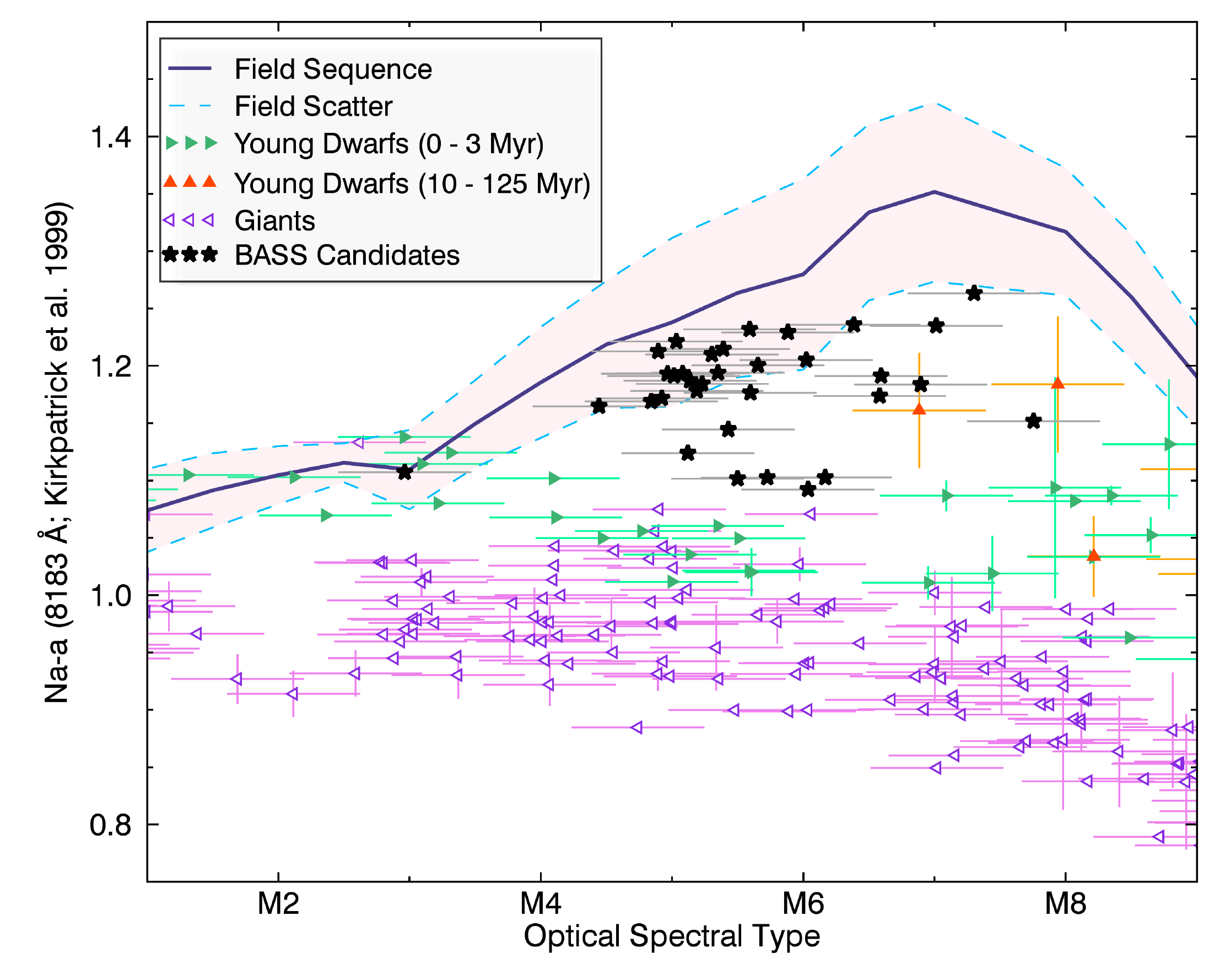}}
	\subfigure[Na-b index ($R \geq 750$)]{\includegraphics[width=0.495\textwidth]{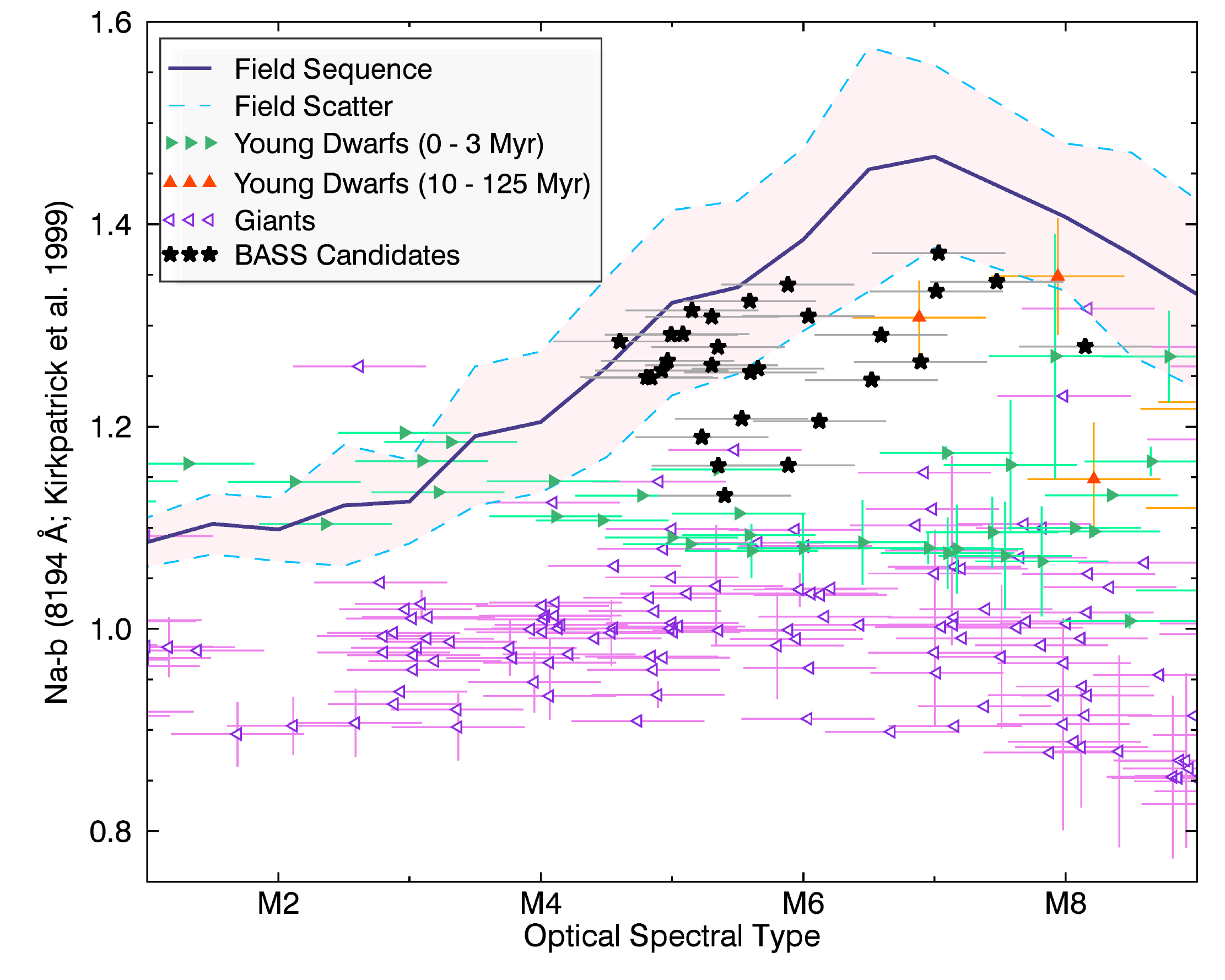}}
	\subfigure[K-a index ($R \geq 750$)]{\includegraphics[width=0.495\textwidth]{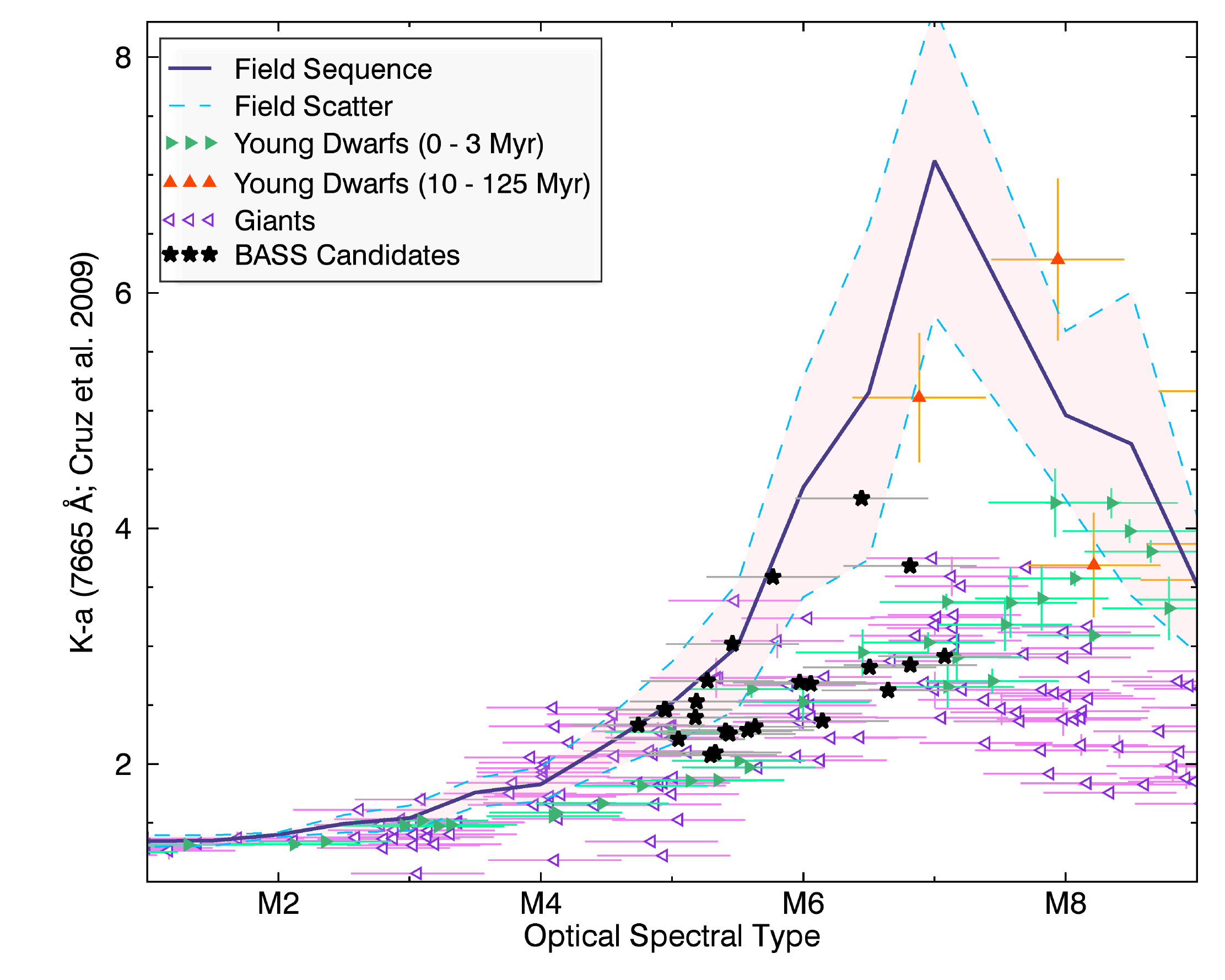}}
	\subfigure[CrH-a index ($R \geq 750$)]{\includegraphics[width=0.495\textwidth]{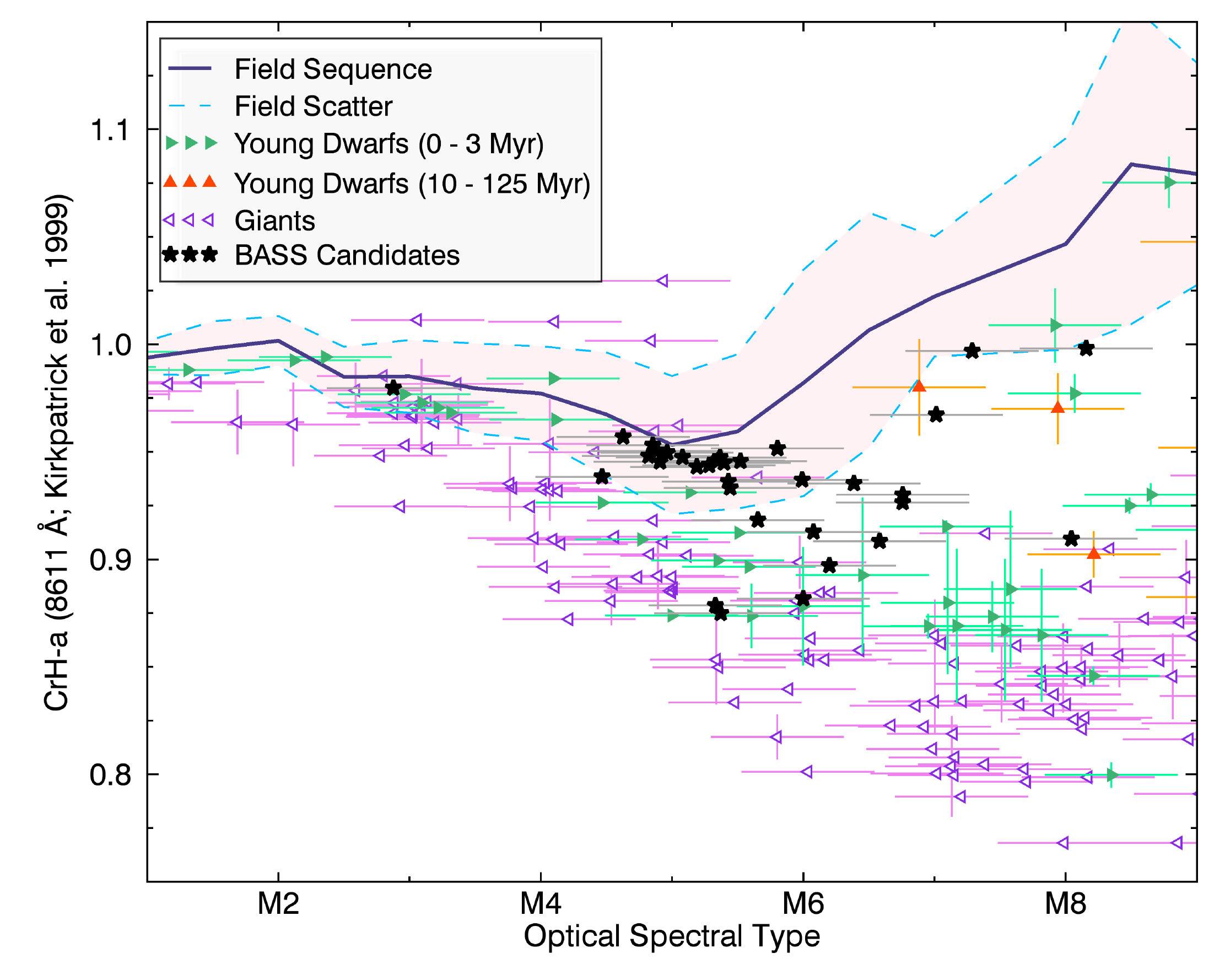}}
	\caption{Gravity-sensitive optical spectroscopic indices of potentially young low-mass stars and BDs in the \emph{BASS} sample (black stars), compared to the field sequence (thick blue line) and its scatter (pink region delimited by dashed pale blue lines). Young members of moving groups considered here (10--125~Myr) are displayed as upside, orange filled triangles, whereas very young objects from star-forming regions (0--3~Myr) are displayed as green, filled right-pointing triangles. Giant stars, which have an even lower surface gravity than those of very young low-mass stars and BDs, are displayed as open, purple left-pointing triangles. It can be seen that giants and young objects follow distinct sequences at spectral types later than M5, however the delimitation is not as clear as for NIR spectroscopic indices (see Figures \ref{fig:I1} and \ref{fig:I2}), especially in the case of early-type young dwarfs.}
	\label{fig:OI}
\end{figure}

In \hyperref[fig:OI]{Figure~\ref*{fig:OI}}, we compare various optical gravity-sensitive indices of M-type dwarfs from \emph{BASS} to preliminary sequences built from field, young and giant M-type stars obtained from the \href{http://dwarfarchives.org}{DwarfArchives}\footnote[2]{\url{http://dwarfarchives.org}}, the Sloan Digital Sky Survey (\emph{SDSS}; \citealp{2009ApJS..182..543A}) and the \href{http://www.astro.umontreal.ca/\textasciitilde gagne/rizzo.php}{Ultracool RIZzo spectral library}\footnote[3]{\url{http://www.astro.umontreal.ca/\textasciitilde gagne/rizzo.php}} (\citealp{2000AJ....120..447K}; \citealp{2002AJ....123.2828C}; \citealp{2003AJ....126.2421C}; \citealp{2007AJ....133..439C}; \citealp{2008AJ....136.1290R}). This allowed us to identify several candidates with signs of low-gravity in the optical, however we find that it is challenging to differentiate young and old $<$ M5 dwarfs using these spectroscopic indices.

%Figure : Allers 2013 Indices, batch 1

%Figure : Allers 2013 Indices, batch 2

%Figure : Optical Indices

\subsection{A New Planetary-Mass Companion to a Young Moving Group Candidate Member}\label{sec:planet}

\begin{figure}
	\centering
	\subfigure[$J$-band image ]{\raisebox{16mm}{\includegraphics[width=0.25\textwidth]{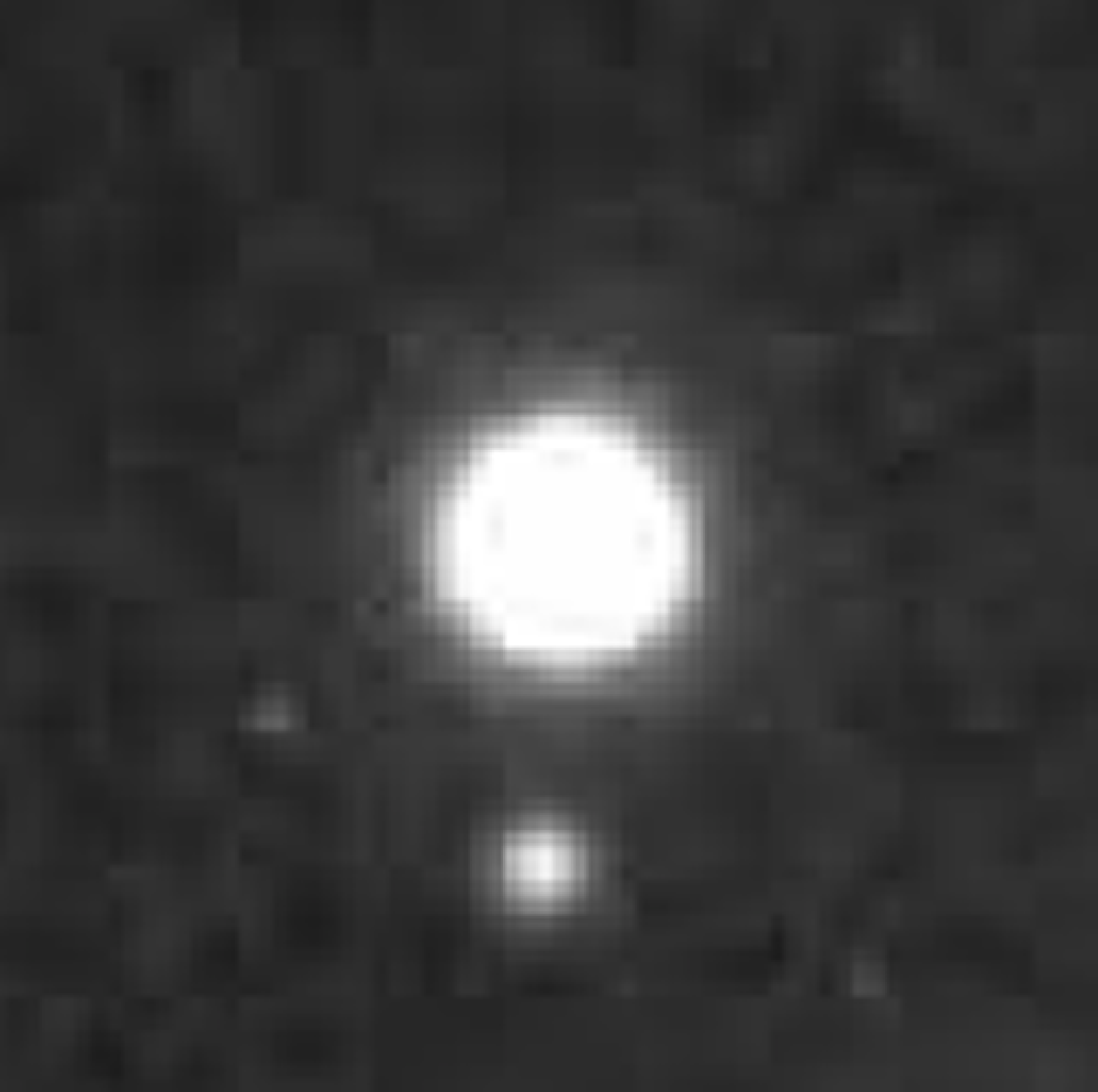}}}
	\subfigure[NIR spectrum]{\includegraphics[width=0.56\textwidth]{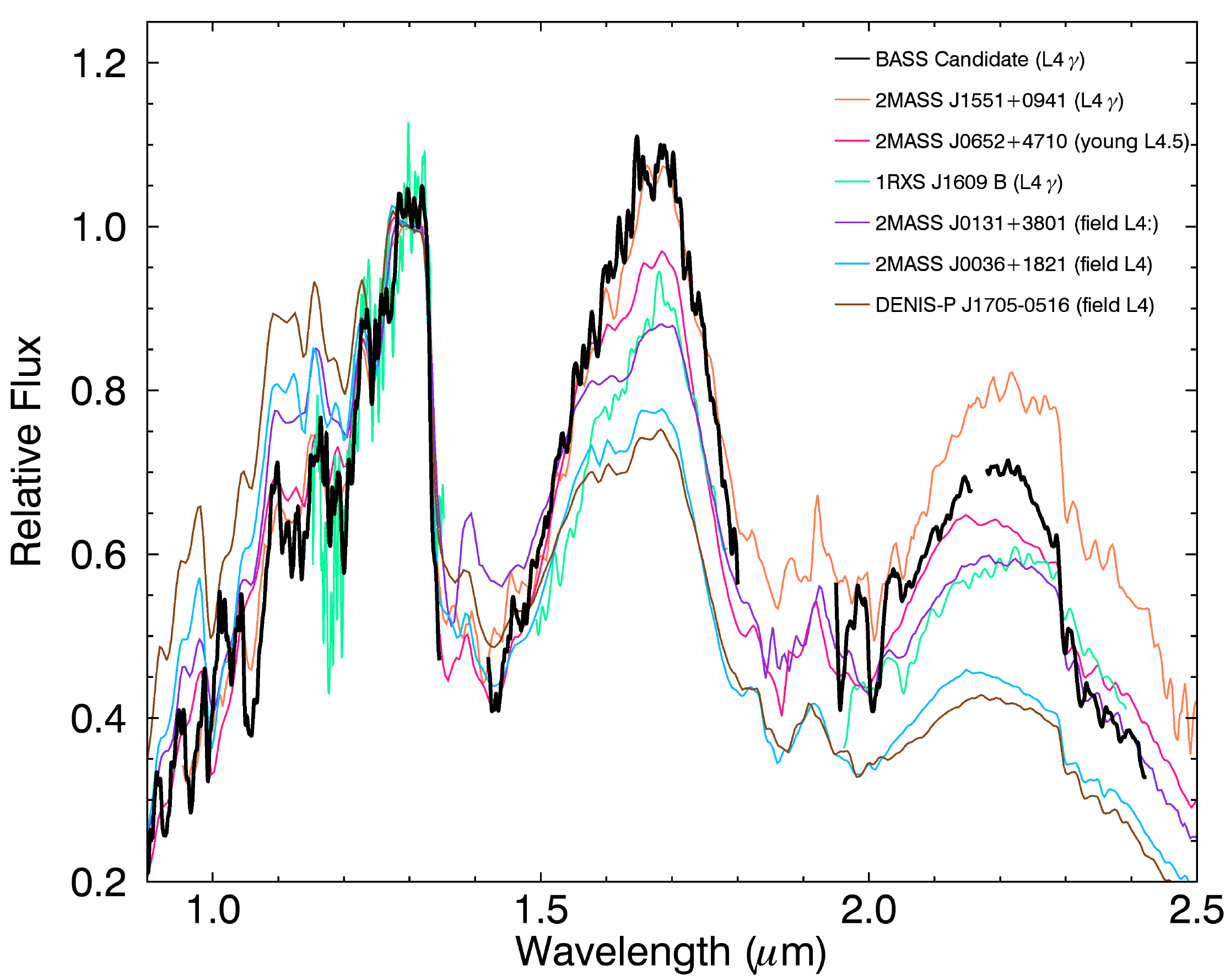}}
	\caption{Left Panel : Direct imaging of a planetary-mass companion to a young low-mass star in the \emph{BASS} sample, using Flamingos-2 in the $J$ band. Right Panel : Resolved FIRE spectrum of the L4~$\gamma$ planetary-mass companion, compared to several field and low-gravity L4 BDs.}
	\label{fig:PLANET}
\end{figure}

We used acquisition images from our spectroscopic follow-up to identify a large-separation co-moving companion to a young low-mass star in the \emph{BASS} sample (\hyperref[fig:PLANET]{Figure~\ref*{fig:PLANET}}). A spectroscopic follow-up with the FIRE spectrometer revealed that the companion is young, with a spectral type L4~$\gamma$. This result will be presented in more detail in an upcoming paper (\href{http://www.astro.umontreal.ca/\textasciitilde gagne/banyanVI.php}{\'E. Artigau et al., in preparation}).

%Figure : Planet

\subsection{Tentative Indications of mass segregation}\label{sec:mseg}

The Virial theorem suggests that a group of isolated and gravitationally interacting stars should relax towards a state of equilibrium where every star possesses the same quantity of kinetic energy. For this reason, it is expected that the massive stars of such a group would have a smaller velocity spread compared to the less massive members. This effect has been observed in several associations of stars (\citealp{2004MNRAS.351.1401J}; \citealp{2011MNRAS.413.2345H}; \citealp{2011A&A...532A.119O}; \citealp{2013ApJ...764...73P}), however it has never been demonstrated for any of the YMGs considered here.\\

\begin{figure}
	\centering
	\includegraphics[width=0.99\textwidth]{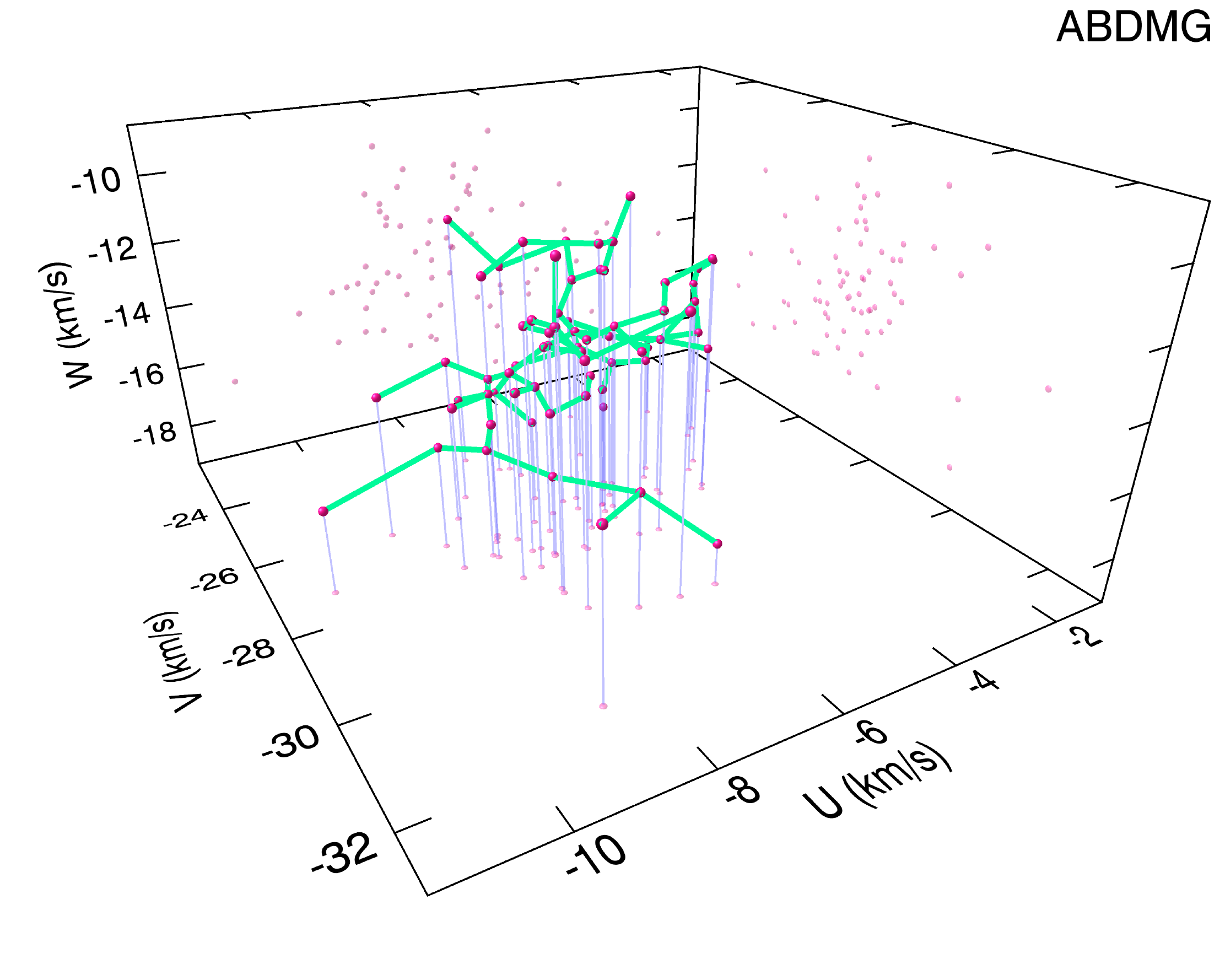}
	\caption{Minimum spanning tree (pale green lines) for bona fide members and \emph{BASS} candidates of ABDMG (red spheres) in $UVW$ velocity space. Red spheres are projected on the three planes for visibility. The minimum spanning tree is the network of minimal length that connects all points together (\emph{i.e.}, all candidates and members of ABDMG), without creating any loop (see text).}
	\label{fig:MST}
\end{figure}

\cite{2011A&A...532A.119O} introduced a quantitative method to measure mass segregation in associations of stars, based on the principle of Minimum Spanning Trees (MSTs). An MST is defined as the shortest network of lines that connects together a set of points, without creating any loop (see \hyperref[fig:MST]{Figure~\ref*{fig:MST}} for an example). This offers the crucial advantage of measuring the typical length scale of a distribution without any prior knowledge on its geometry. For example, determining the MST of the spatial $XYZ$ distribution of members of a YMG will yield its length scale without needing to determine its center of mass. This property is crucial in the case of YMGs, because we expect many of their members to be still missing, not to mention that the masses of those that we know are generally not well constrained. \\

%Figure : MST

We use a metric for the measurement of mass segregation that is defined by \cite{2009MNRAS.395.1449A}; the total length of the MST is determined for a subset of the $N$ most massive stars of a given association (we call this measurement $l_{\mathrm{massive}}$), and then this is repeated for a large number of random subsets of $N$ stars in the same association (which yields an array of lengths $l_{\mathrm{norm};i}$). The Mass Segregation Ratio (MSR; not to be confused with MST) is then defined as :

\begin{equation}
	\Lambda_{\mathrm{MSR}} = \frac{<l_{\mathrm{norm};i}>}{l_{\mathrm{massive}}},
\end{equation}

where $< x_i >$ denotes the mean value of an array $x_i$. The standard deviation of $l_{\mathrm{norm};i}$ can be used in turn to assess the statistical signification of $\Lambda_{\mathrm{MSR}}$. \\

\begin{figure}
	\centering
	\subfigure[Dynamical MSR of ABDMG]{\includegraphics[width=0.495\textwidth]{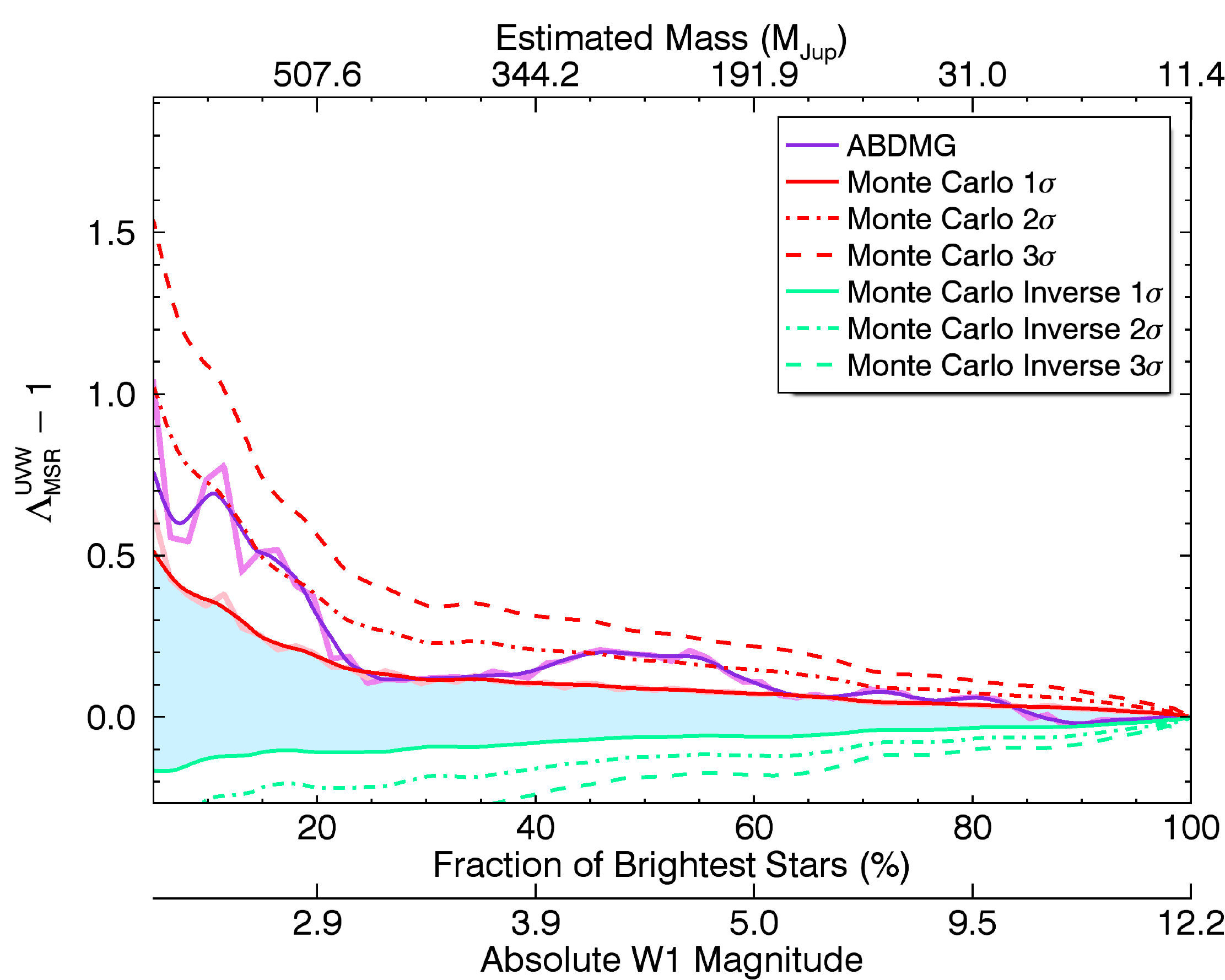}}
	\subfigure[Spatial MSR of ARG]{\includegraphics[width=0.495\textwidth]{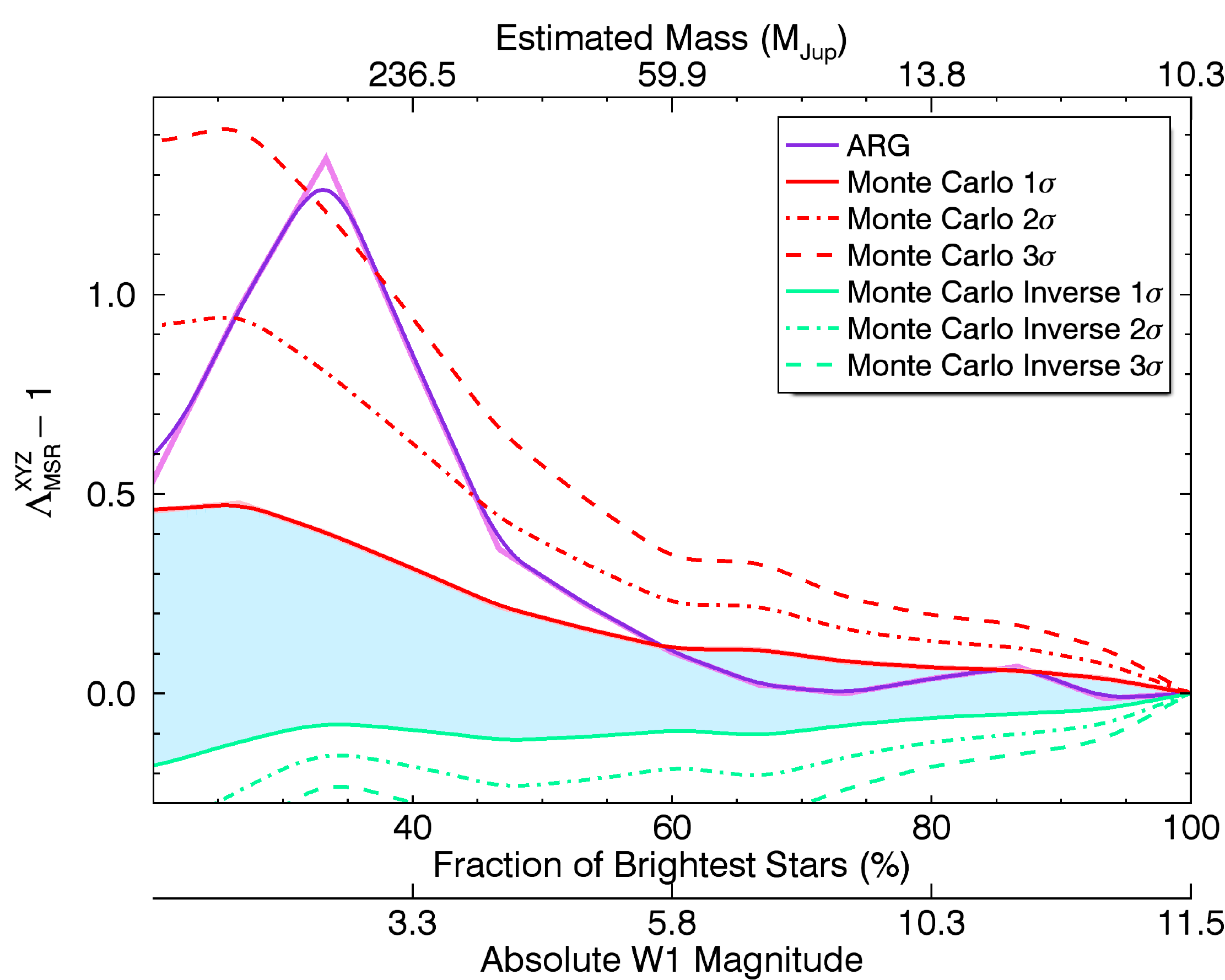}}
	\caption{Dynamical ($UVW$ space) and spatial ($XYZ$ space) mass segregation ratios ($\Lambda_{\mathrm{MSR}}$) for all bona fide members and \emph{BASS} candidates in ABDMG and ARG\index{Argus} (pale purple line; dark purple line is a smoothed version). A unit MSR indicates no mass segregation (\emph{i.e.}, the MST length does not correlate with the absolute magnitude of the subset of objects used to define the MSR). An MSR larger (smaller) than one indicates that the brightest (faintest) stars are more concentrated in position or velocity space. A Monte Carlo simulation allowed us to delimit the region where $\Lambda_{\mathrm{MSR}}$ does not depart from unity with a statistical significance of 1$\sigma$ (pale blue region). Red lines indicate a positive departure from $\Lambda_{\mathrm{MSR}} = 1$ at significances of 1$\sigma$ (solid line), 2$\sigma$ (dash-dotted line) and 3$\sigma$ (dashed line), whereas green lines similarly indicate a negative departure from $\Lambda_{\mathrm{MSR}} = 1$. For example, panel (b) indicates that an MST built from only the 30\% brightest stars in ARG\index{Argus} (corresponding to $M_{W1} \leq 2.7$ or estimated masses $\geq 0.27$ \Msol), has a larger total length than that of an MST build from random selections of 30\% of stars in ARG\index{Argus}, at a $> 3\sigma$ significance. This indicates that the stars in this sample which are more massive than $\sim 0.27$ \Msol\ tend to be more concentrated in the $XYZ$ space.}
	\label{fig:SEG}
\end{figure}

We applied this method to the spatial $XYZ$ and kinematic $UVW$ distributions of bona fide members and \emph{BASS} candidates of YMGs considered here, for all possible values of $N$. Instead of selecting subsets of stars based on their estimated masses which are uncertain, we rather use their absolute $M_{W1}$ magnitudes, which should vary monotonically with mass for a coeval population. We present partial results in \hyperref[fig:SEG]{Figure~\ref*{fig:SEG}}: we find that objects in ARG\index{Argus} have an MSR larger than unity at a 3$\sigma$ significance when selecting the subset of stars more massive than $\sim$ 0.27 \Msol\ (corresponding to a positive mass segregation, \emph{i.e.} more massive stars being more concentrated), whereas we find positive dynamical mass segregation in ABDMG with a $\sim$ 1$\sigma$ significance if we select stars more massive than $\sim$ 0.48 \Msol\ or $\sim$ 0.22 \Msol. More detail on these results will be presented in \href{http://www.astro.umontreal.ca/\textasciitilde gagne/banyanV.php}{J.~Gagn\'e et al. (submitted to ApJ)}, but we point out that measurements of radial velocity and parallax for all \emph{BASS} candidates will be needed to corroborate these tentative results.

%Figure : SEG RATIO

\section{CONCLUSIONS}

We present a description of the candidate selection method that we used to build the \emph{BASS} catalog of very low-mass candidate members of YMGs, as well as a spectroscopic follow-up to identify signs of youth in their NIR and optical spectra. We present first results from this survey, including several new low-mass stars and BDs displaying telltale signs of youth such as a triangular $H$-band continuum or lower-than-normal atomic line equivalent widths. We adapt the method of minimum spanning trees to YMGs, and use it to identify tentative signs of spatial and dynamical mass segregation in YMGs, however we stress that a complete measurement of the kinematics of the \emph{BASS} sample will be needed to verify these results.

\acknowledgments{
The authors would like to thank Kelle Cruz, Katelyn Allers, David Rodriguez, Philippe Delorme, Micka\"el Bonnefoy, Andr\'e-Nicolas Chen\'e, Adric Riedel, Lo\"ic Albert, Ben Oppenheimer, David Blank, Am\'elie Simon, Carlo Felice Manara, Jonathan Foster and Zahed Wahhaj for useful comments and discussions and/or for sharing data. We also wish to thank various telescope operators and staff scientists which were of great help through our numerous observing runs; Stuart Ryder, Rub\'en Diaz, St\'epanie C\^ot\'e, Mischa Schirmer, John Blakeslee, Sandy Leggett, Mike Connelley, Brian Cabreira, Bill Golisch, Dave Griep and Tony Matulonis. This work was supported in part through grants from the the Fond de Recherche Qu\'eb\'ecois - Nature et Technologie and the Natural Science and Engineering Research Council of Canada. This research has benefitted from the SpeX Prism Spectral Libraries, maintained by Adam Burgasser at \url{http://pono.ucsd.edu/\textasciitilde adam/browndwarfs/spexprism}, and the \emph{Database of Ultracool Parallaxes} at \url{http://www.cfa.harvard.edu/\textasciitilde tdupuy/plx/Database\_of\_\\ Ultracool\_Parallaxes.html}. This research made use of; the SIMBAD database and VizieR catalog access tools, operated at Centre de Donn\'ees astronomiques de Strasbourg, France \citep{2000A&AS..143...23O}; data products from the Two Micron All Sky Survey (\emph{2MASS}), which is a joint project of the University of Massachusetts and the Infrared Processing and Analysis Center (IPAC)/California Institute of Technology (Caltech), funded by the National Aeronautics and Space Administration (NASA) and the National Science Foundation (NSF; \citealp{2006AJ....131.1163S}); data products from the Wide-field Infrared Survey Explorer (\emph{WISE}), which is a joint project of the University of California, Los Angeles, and the Jet Propulsion Laboratory (JPL)/Caltech, funded by NASA \citep{2010AJ....140.1868W}; the NASA/IPAC Infrared Science Archive, which is operated by the JPL, Caltech, under contract with NASA; the M, L, and T dwarf compendium housed at \url{http://DwarfArchives.org} and maintained by Chris Gelino, Davy Kirkpatrick, and Adam Burgasser. This work is based on observations obtained at the Gemini Observatory (program numbers GS-2012B-Q-70, GN-2013A-Q-106, GS-2013A-Q-66, GS-2013B-Q-79, GN-2013B-Q-85, GS-2014A-Q-55 and GN-2014A-Q-94), which is operated by the Association of Universities for Research in Astronomy, Inc., under a cooperative agreement with the NSF on behalf of the Gemini partnership: the NSF (United States), the National Research Council (Canada), the Comisi\'{o}n Nacional de Investigaci\'{o}n Cient\'{i}fica y Tecnol\'{o}gica (Chile), the Australian Research Council (Australia), Minist\'{e}rio da Ci\^{e}ncia, Tecnologia e Inova\c{c}\~{a}o (Brazil) and Ministerio de Ciencia, Tecnolog\'{i}a e Innovaci\'{o}n Productiva (Argentina). This work is based on observations obtained with the Flamingos~2 spectrometer, which was designed and constructed by the IR instrumentation group (PI: R. Elston) at the University of Florida, Department of Astronomy, with support from NSF grant AST97-31180 and Kitt Peak National Observatory. This paper includes data gathered with the 6.5 meter Magellan Telescopes located at Las Campanas Observatory, Chile (CNTAC program CN2013A-135). This research is based on observations from the Infrared Telescope Facility (program numbers 2012A097, 2012B015, 2013A055 and 2013B025), which is operated by the University of Hawaii under Cooperative Agreement no. NNX-08AE38A with the NASA, Science Mission Directorate, Planetary Astronomy Program. The authors recognize and acknowledge the very significant cultural role and reverence that the summit of Mauna Kea has always had within the indigenous Hawaiian community. We are most fortunate to have the opportunity to conduct observations from this mountain.
}

\indent \emph{Facilities:} IRTF (SpeX), Magellan:Baade (FIRE), Gemini:South (Flamingos~2, GMOS), Gemini:North (GNIRS, GMOS).

\end{document}